\documentclass[journal,twocolumn]{IEEEtran}

\usepackage{cite} 
\usepackage{xspace}
\usepackage{filecontents}
\usepackage{float}
\usepackage{pgf, tikz}
\usepackage{graphicx}
\usepackage[cmex10]{amsmath}
\usepackage{mathtools, cuted}

\usepackage{bm}
\usepackage{amssymb} 
\usepackage{array}
\usepackage{amsfonts}
\usepackage{amsthm}   
\usepackage{enumerate} 
\usepackage{enumitem}
\usepackage{multirow}
\usepackage{multicol}
\usepackage{cleveref}
\usepackage{lipsum}
\crefrangeformat{equation}{(#3#1#4)\,--\,(#5#2#6)}

\newtheorem{proposition}{Proposition}
\newtheorem{remark}{Remark}
\usetikzlibrary{patterns,snakes}

\usepackage{stfloats,epstopdf}
\usepackage{array,algorithm,algorithmic}
\usepackage{enumerate}
\usepackage{pgfplots}
\usepackage{caption}
\usepackage{subcaption}
\usepackage{tabularx}
\usepackage{pgfplotstable}
\usepgfplotslibrary{dateplot}
\usepackage{cite}
\usepackage{url}
\usepackage{amsbsy}
\usepgfplotslibrary{patchplots}
\usepackage{xcolor}
\usepackage{tikz}
\usetikzlibrary{shadows}
\usetikzlibrary{fadings}
\usetikzlibrary{arrows}
\usetikzlibrary{shapes,snakes}
\pgfplotsset{compat=1.7}
\usetikzlibrary{patterns}
\usetikzlibrary{spy}

\usepackage{amsmath,accents}
\DeclareMathSymbol{\widehatsym}{\mathord}{largesymbols}{"62}
\newcommand{\sugcom}[1]{{\color{black}#1}\xspace}
\newcommand{\redcom}[1]{{\color{black}#1}\xspace}

\DeclareMathSymbol{\widetildesym}{\mathord}{largesymbols}{"65}

\pgfplotscreateplotcyclelist{mycyclelist}{
solid, red, line width=1.0pt, every mark/.append style={solid, fill=red}, mark=square, mark size={1.2}\\%
densely dashed, blue, line width=1.0pt, every mark/.append style={solid, fill=blue}, mark=triangle, mark size={1.2}\\%
densely dotted, line width=1.0pt, green!70!black, every mark/.append style={solid, fill=green!70!black}, mark=o, mark size={1.2}\\%
densely dashdotted, line width=1.0pt, brown!70!black, every mark/.append style={solid, fill=brown!70!black}, mark=diamond, mark size={1}\\%
dotted, line width=1.0pt, orange!70!black, every mark/.append style={solid, fill=orange!70!black}, mark=x, mark size={1.3}\\%
thick, solid,  red\\%
densely dashdotted, line width=1.0pt, blue\\%
densely dotted,  line width=1.2pt, green!70!black\\%
solid,  blue, every mark/.append style={solid, fill=blue}, mark=*, mark size={1}\\%
dashed, blue, every mark/.append style={solid, fill=blue}, mark=*, mark size={1}\\%
solid,  red, every mark/.append style={solid, fill=red}, mark=square*, mark size={1}\\%
dashed, red, every mark/.append style={solid, fill=red}, mark=square*, mark size={1}\\%
solid,  green!60!black, every mark/.append style={solid, fill=green!60!black}, mark=triangle*, mark size={1.2}\\%
dashed, green!60!black, every mark/.append style={solid, fill=green!60!black}, mark=triangle*, mark size={1.2}\\%
solid,  brown!70!black, every mark/.append style={solid, fill=brown!70!black}, mark=diamond*, mark size={1.2}\\%
dashed, brown!70!black, every mark/.append style={solid, fill=brown!70!black}, mark=diamond*, mark size={1.2}\\%
}
\pgfplotscreateplotcyclelist{mycyclelist2}{
densely dashed, blue, line width=1.0pt, every mark/.append style={solid, fill=blue}, mark=triangle, mark size={1.2}\\%
densely dashdotted, line width=1.0pt, red, every mark/.append style={solid, fill=brown!70!black}, mark=x, mark size={1.2}\\%
densely dotted, line width=1.0pt, green!70!black, every mark/.append style={solid, fill=green!70!black}, mark=o, mark size={1.2}\\%
thick, solid, blue, line width=1.0pt, every mark/.append style={solid, fill=blue}, mark=triangle, mark size={1.2}\\%
thick, solid, line width=1.0pt, red, every mark/.append style={solid, fill=brown!70!black}, mark=x, mark size={1.2}\\%
thick, solid, line width=1.0pt, green!70!black, every mark/.append style={solid, fill=green!70!black}, mark=o, mark size={1.2}\\%
thick, solid,  red\\%
densely dashdotted, line width=1.0pt, blue\\%
thick, solid,  green!70!black\\%
dashed, line width=1.0pt, brown!70!black\\
densely dashdotted, blue, every mark/.append style={solid, fill=blue}, mark=square, mark size={1}\\%
solid, blue, every mark/.append style={solid, fill=blue}, mark=triangle, mark size={1.2}\\%
densely dotted, line width=0.5pt, blue, every mark/.append style={solid, fill=blue}, mark=o, mark size={1}\\%
dashed, blue, every mark/.append style={solid, fill=blue}, mark=x, mark size={1.2}\\%
thick, dashed,  green!70!black\\%
densely dashdotted, red, every mark/.append style={solid, fill=red}, mark=square, mark size={1}\\%
solid, red, every mark/.append style={solid, fill=red}, mark=triangle, mark size={1.2}\\%
densely dotted, red, every mark/.append style={solid, fill=red}, mark=o, mark size={1}\\%
dashed, red, every mark/.append style={solid, fill=red}, mark=x, mark size={1.2}\\%
}
\pgfplotscreateplotcyclelist{mycyclelist3}{
solid,  blue, every mark/.append style={solid, fill=blue}, mark=*, mark size={1}\\%
densely dashed, red, every mark/.append style={solid, fill=red}, mark=x, mark size={1.5}\\%
}

\newcommand{\AuthorOne}{Sahar~Idrees, {\em{Student Member, IEEE}}}
\newcommand{\AuthorTwo}{Xiangyun~Zhou, {\em{Senior Member, IEEE}}}
\newcommand{\AuthorThree}{Salman~Durrani, {\em{Senior Member, IEEE}}}
\newcommand{\AuthorFour}{Dusit~Niyato, {\em{Fellow, IEEE}}}

\newcommand{\ThankOne}{

}
\newcommand{\ThankTwo}{
S. Idrees, X. Zhou and S. Durrani are with the Research School of Electrical, Energy and Materials Engineering, College of Engineering and Computer Science, Australian National University (ANU), Canberra, ACT 2601, Australia (Emails: \{sahar.idrees, xiangyun.zhou, salman.durrani\}@anu.edu.au). S. Idrees is currently on study leave from UET, Lahore pursuing her PhD at ANU, Canberra. D. Niyato is with the School of Computer Science and Engineering, Nanyang Technological University, 50 Nanyang Ave, Singapore 639798 (email: dniyato@ntu.edu.sg).
}
\newcommand{\ThankThree}{This work was supported by the Australian Research Councils Discovery Project Funding Scheme under Project DP170100939.

}
\newcommand{\ThankFour}{
Part of this work has been accepted for presentation in IEEE ICC 2020~\cite{Idrees-ICC2020}.
}

\begin{document}

\title{Design of Ambient Backscatter Training for Wireless Power Transfer}
\author{\IEEEauthorblockN{\AuthorOne,~\AuthorTwo, \AuthorThree, and~\AuthorFour\thanks{\ThankOne}\thanks{\ThankTwo}\thanks{\ThankThree}\thanks{\ThankFour}}}
\maketitle


\setcounter{page}{1}
\begin{abstract}
Wireless power transfer (WPT) using energy beamforming is a promising solution for low power Internet of Things (IoT) devices. In this work, we consider WPT from an energy transmitter (ET) employing retrodirective WPT using a large phased antenna array to an energy receiver (ER) capable of ambient backscatter. The advantage of retrodirective WPT is that no explicit channel estimation is needed at the ET and the use of ambient backscattering eliminates the need for active transmission at the ER. We propose a training sequence design, i.e., pattern of varying the reflection coefficient at the ER, to eliminate the direct-link interference from the ambient source.  We show that when the ambient symbol duration is known, the ambient interference is fully cancelled by the proposed design. We analytically model the system and find the average harvested power at the ER considering Nakagami-$m$ fading channels and non-linear energy harvesting model. Our results clearly show that the proposed solution is robust to a small timing offset mismatch at the correlator. When interference from undesired neighbouring sources in the ambient environment is not significant, the ER can successfully harvest tens to hundreds of $\mu$W of power, which is an important improvement for low-power IoT devices.
\end{abstract}
\IEEEpeerreviewmaketitle

\begin{IEEEkeywords}
Ambient backscatter communication, wireless power transfer, training sequence design, direct sequence spread spectrum.
\end{IEEEkeywords}

\section{Introduction}\label{intro}
\subsection{Motivation and Related Work}
The Internet of Things (IoT) is currently making a rapid transition from theory to practice. For instance, in Australia large scale IoT networks targeting smart cities~\cite{smartcity1},~\cite{smartcity2} and smart agriculture~\cite{agriculture} are currently being deployed. As we move towards a world filled with a large number of IoT devices, the means to sustainably powering these IoT devices is a key challenge. In this regard, far-field wireless power transfer (WPT) is a promising technology to provide convenient wireless charging to low power IoT devices~\cite{jayakody2017wireless,huang2015cutting,Bi2015,zeng2017sigdes}.\\
\indent The problem of efficient WPT from an energy transmitter (ET) to an energy receiver (ER) has received much attention in the literature~\cite{Zhang-2018survey,Alsaba2018,Liu-Zhang2014,yZeng2015a,yZeng2015b,gYang2014,xChen2014,hSon2014,park2015joint,jXu-2014}. Typically, WPT relies on highly directional beamforming to increase the end-to-end power efficiency and overcome the severe radio frequency (RF) signal attenuation over distance. In this regard, different beamforming architectures have been proposed~\cite{Zhang-2018survey,Alsaba2018,Liu-Zhang2014}. However, the implementation of these beamforming architectures requires channel state information (CSI) estimation at the ET~\cite{yZeng2015a,yZeng2015b} or at the ER~\cite{gYang2014,xChen2014,hSon2014,park2015joint}, or energy feedback from the ER to the ET~\cite{jXu-2014,Zeng-2015}. CSI estimation at the ER increases the complexity of the ER, which is undesirable. In addition, training methods suffer from high feedback overhead, which should also be avoided.\\
\indent \sugcom{Employing the concept of retrodirectivity is a promising solution to avoid the need for any CSI estimation for efficient WPT~\cite{zeng2017sigdes,Massa-2013}. Originally, retrodirective arrays, such as Van Atta array~\cite{Vanatta-1960} and Pon array~\cite{Pon-1964}, were proposed as `reflection type' arrays to reflect an incident signal back to the direction that it came from. This reflection of incoming waves is realized by their reversal in the time domain or phase conjugation in the frequency domain. Recently, more advanced versions of arrays employing the retrodirective principle have been developed for the purpose of WPT~\cite{Massa-2013,Jandhyala-2012,RodenbeckPhased-2004,Hsieh-2003,RodenbeckLimitation-2005}. The retrodirective WPT exploits channel reciprocity and provides WPT without explicit channel estimation. In particular, it involves the ET equipped with a phased array, providing WPT to an ER. This is accomplished by the ET first receiving a signal from the prospective ER, which then serves as a reference signal to steer a beam back towards the ER. This is done by conjugating this received signal and using this conjugated signal to set the phase of an energy signal such that it is emanated towards the ER~\cite{zeng2017sigdes}. In this regard, a novel massive MIMO retrodirective WPT scheme was proposed in~\cite{Lee-2018}. However, this scheme still required active signal transmission from the ER to initiate WPT, which consumes energy and may not be desirable for low power IoT devices. The active signal transmission from the ER to the ET was avoided in~\cite{Yang-2015csi} by enabling the ER to backscatter the pilots emitted by the ET. However, conventional beamforming was still employed at the ET. A WPT scheme employing monostatic backscatter at the ER and retrodirective WPT at the ET was proposed in~\cite{Krikidis-2018}. However, the charging request was initiated by the ET using active transmission.}\\
\indent Backscatter communication is a promising ultra-low power wireless communication paradigm, which eliminates the need for active transmission by the low power IoT devices~\cite{Huynh-2018,liu2019next}. Conventional monostatic backscatter systems enable a tag to transmit to the reader by reflecting the RF signal sent by the reader itself. Recently, ambient backscatter which enables the tag to make use of ambient RF signals generated from ambient RF sources for communication has attracted a lot of attention~\cite{DevineniBER-2019,liu-2013ambient,parks2015turbocharging,Kimionis-2014,Wang-2016,Qian-2017,DevineniNonCoh-2019,kellogg2016passive,Yang-2018air,zhang2016enabling,iyer2016inter,bharadia2015backfi,zhang2016hitchhike}. A key issue in ambient backscatter communication systems is the direct-link interference that the RF ambient source causes to the tag. This is due to the fact that the ambient signals are omnipresent and much stronger than their backscattered versions. Numerous works in literature \sugcom{evaluate the impact of this direct-link interference on different aspects of system performance, e.g., bit error rate (BER) of ambient backscatter communication~\cite{DevineniBER-2019}} and propose different techniques to resolve this issue~\cite{kellogg2016passive,Kimionis-2014,Wang-2016,Qian-2017,DevineniNonCoh-2019,zhang2016enabling,iyer2016inter,Yang-2018air,bharadia2015backfi,zhang2016hitchhike}. One approach is to consider this direct-link interference as a component of the background noise~\cite{Kimionis-2014,Wang-2016,Qian-2017}. However, since the backscatter signal is very weak as compared to the ambient signal, such schemes do not perform so well. \sugcom{~\cite{DevineniNonCoh-2019} demonstrated the existence of a BER floor in a single antenna backscatter device and used multiple antennas to cancel the direct link interference in a non-coherent receiver setup.} Other approaches involve general signal processing techniques~\cite{kellogg2016passive,Yang-2018air,zhang2016enabling,iyer2016inter} or backscatter specific solutions such as frequency shifting~\cite{bharadia2015backfi,zhang2016hitchhike}. In this regard, to the best of our knowledge, the use of Direct Sequence Spread Spectrum (DSSS) has not been considered to date.
\subsection{Our Contributions}
\sugcom{In this paper, we consider a scenario with an ET equipped with a large phased antenna array capable of retrodirective WPT and an ER equipped with an ambient backscatter tag. The fundamental signal recovery problem at the ET is then: \textit{How to recover the weak backscattered signal in the presence of strong direct-link ambient interference?} We consider this problem assuming general Nakagami-$m$ fading and non-linear energy harvesting model.} In this context, our main contributions are:
\begin{itemize}
\item Taking inspiration from DSSS, we consider an ambient backscatter training scheme in which we vary the backscatter coefficient at the ER. This in effect multiplies the backscattered signal with a DSSS training signal and aims to capitalize on the spreading gain to boost the backscattered signal. We show that with a pseudo-noise (PN) training sequence, the average harvested power at the ER is small and it even reduces as the training period increases. This is due to the fact that the use of PN training sequence completely fails in dealing with the strong direct-link ambient interference.
\item We then propose the design of the training sequence (i.e., the pattern of varying the reflection coefficient), to completely eliminate the direct-link ambient interference. We show that when the ambient symbol duration is known, the ambient interference is cancelled as long as there are equal number of $+1$ and $-1$ chips over one ambient symbol. The number of chips or equivalently the switching rate does not matter in this case. Hence, we can use the slowest switching rate, i.e., we can switch the backscatter coefficient only twice per ambient symbol period. We analytically model the system and derive a closed-form expression for the average harvested power at the ER. We show that this deterministic training sequence scheme has superior performance as compared to the PN training sequence scheme.
\item Finally, we show that the proposed solution is robust to small timing offset mismatch at the correlator. This is because the undesired component is still perfectly eliminated. However, good synchronization is needed for the best performance. In addition, when the ambient duration is unknown, the power transfer performance under the proposed deterministic training scheme can be severely degraded. This is due to unequal durations of $+1$ and $-1$ chips in one ambient symbol. We show that in this mismatched case, the number of chips does matter, i.e., it is best to use a fast switching rate to minimize the effect of the uncancelled ambient. \redcom{In addition, we consider interference from neighbouring signals in the ambient environment, which is shown to impact the energy harvesting performance. However, the system can still harvest tens to hundreds of $\mu$W of power if these interference signals from neighbouring ambient sources are significantly weaker than the direct-link ambient signal.}
\end{itemize}
\begin{table}[t]
\centering
\caption{Summary of main mathematical symbols.}\label{tb:2}
\begin{tabular}{|c||c|l|} \hline
 & Symbol & Description \\ \hline%
\parbox[t]{1mm}{\multirow{20}{*}{\rotatebox[origin=c]{90}{System Parameters}}}
 &&\\
  &$\alpha$& Path-loss exponent \\
  &$\gamma$& Large scale channel attenuation \\
  &$\beta$& Backscatter coefficient   \\
  &${m_g}$& Nakagami fading parameter for AS $\,\to\,$ ER link \\
  &${m_h}$ & Nakagami fading parameter for AS $\,\to\,$ ET link\\
  &${m_f}$& Nakagami fading parameter for ER $\,\to\,$ ET link \\
  &$\sigma_n^2$& Variance of AWGN \\
  &  $d_1$& Distance between the AS and the ER \\
  &$d_2$& Distance between the ER and the ET  \\
  &$d_3$& Distance between the AS and the ET\\
  &$P_s$&  Transmit power of the AS\\
  & ${T_b}$ & Duration of backscatter phase \\
  & ${T_c}$ & Chip duration (fixed backscatter coefficient)\\
  & ${T_s}$ & Duration of one ambient symbol \\
  &  ${T_{\textrm{off}}}$ & Duration of offset mismatch at the correlator\\
  & $N_c$& Number of chips during backscatter phase  \\
  & $M$ &  Number of antennas at the ET \\
  & $N_s$  & Number of ambient signals in one backscatter phase \\
  & ${c_n}$ & $n$-th chip in the training sequence \\
   &  $P_t$&  Transmit power of the ET \\
   &&\\
  \hline
\parbox[t]{1mm}{\multirow{8}{*}{\rotatebox[origin=c]{90}{Random Variables}}}
&&\\
 &  ${s_i}$ & $i$-th ambient symbol  \\
 & $g$ & Channel from the AS to the ER  \\
 & $\mathbf{h}$ & Channel from the AS to the ET \\
 & $\mathbf{f}$ & Channel from the ER to the ET\\
 & $\mathbf{{r}_{\textrm{ET}}}$ & Signal received at the ET during the backscatter phase\\
 & $\mathbf{{r}_{\textrm{ER}}}$ &Signal received at the ER during PT phase  \\
 &&\\
 \hline
\end{tabular}
\end{table}
\subsection{Notation and Paper Organization}
The following notation is used in this paper. Pr($\cdot$) indicates the probability measure and $\mathop{\mathbb{E}}[\cdot]$ denotes the expectation operator. $f_X (x)$ denotes the probability density function (pdf) of a random variable $X$. For a complex valued vector $\mathbf{v}$, $\mathbf{v}^*$, $\mathbf{v}^T$ and $\mathbf{v}^H$ denote the conjugate, transpose and conjugate transpose, while the norm of the vector $\mathbf{v}$ is given by $\left\|\mathbf{v}\right\| = \sqrt{\mathbf{v}^T \mathbf{v}}$. \sugcom{Finally, $\exp(\cdot)$ is the exponential function.} A list of the main mathematical symbols employed in this paper is given in Table~\ref{tb:2}.\\
\indent \sugcom{The rest of the paper is organized as follows. Section~\ref{sys_model} describes the system model and assumptions, along with the proposed wireless power transfer scheme and its phases. Section~\ref{sig_model} presents the signal model of the system in terms of mathematical equations and defines the metric of interest. Section~\ref{PNseq} gives the analysis of the proposed scheme with a PN sequence applied at the ER. Section~\ref{detseq} proposes the design of the deterministic training sequence for the elimination of the direct-link ambient interference and also gives the analysis of the system in this scenario. Section~\ref{imperf} deals with the impact of practical system aspects like imperfect synchronization at the correlator and change in ambient symbol duration. Section~\ref{res} presents the numerical results. Finally, Section VI concludes the paper}.
\section{System Model}\label{sys_model}
\begin{figure}
\centering
\includegraphics[width=0.5  \textwidth]{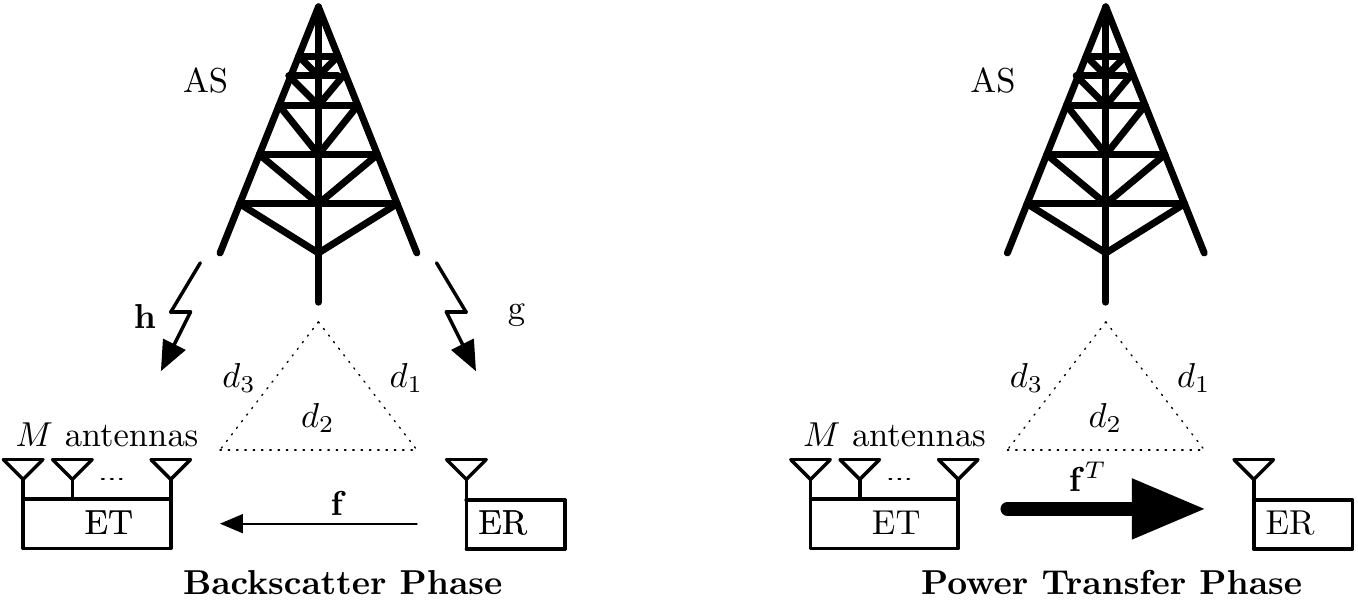}
        \caption{Illustration of the system model.}
        \label{smodel}
\end{figure}

\indent We consider a WPT scenario with an ambient source (AS), an energy transmitter (ET) and an energy receiver (ER). The signal broadcasted from the AS is received by both the ET and the ER. We study the design of wireless power transfer (WPT) from the ET to the ER, as illustrated in Fig.~\ref{smodel}.\\
\indent The ER is a device (e.g., a sensor) that is capable of backscatter transmissions. It is composed of a single antenna element, a micro-controller, a variable impedance and an energy harvester. We also assume that the ER is equipped with an ideal energy storage element (e.g., a supercapacitor) for storing the energy transferred by the ET. The block diagram of the ER is illustrated in Fig.~\ref{bdiag}a.\\
\indent The ET is connected to the power grid and transmits with a fixed power $P_t$ using a \sugcom{phased antenna array} with $M$ elements where $M$ is large, which ensures that the ET forms a thin focussed beam. The block diagram of the ET is illustrated in Fig.~\ref{bdiag}b.
\subsection{Channel Assumptions}
\indent We assume that all the channel links are composed of large-scale path loss, with exponent $\alpha$. \sugcom{The block fading for all links is modelled as the independent and identically  distributed  (i.i.d.)  Nakagami-$m$ fading with $m_g$, $m_f$ and $m_h$ being the Nakagami-$m$ parameters of the AS to ER, ER to ET and AS to ET channels respectively.} We denote the distances between AS $\,\to\,$ ER, ER $\,\to\,$ ET and AS $\,\to\,$ ET by $d_1$, $d_2$ and $d_3$ respectively. Thus, large-scale attenuation is modelled as $ \gamma_i = k_0 (d_i/d_0)^{-\alpha} $ where $k_0$ is the constant attenuation for path-loss at a reference distance of $d_0$ and $i \in \{1,2,3\}$.\\
\indent The ER $\,\to\,$ ET, AS $\,\to\,$ ET and AS $\,\to\,$ ER   fading channel coefficients, denoted by $\mathbf{f}$, $\mathbf{h}$ and $g$ respectively, are modeled as quasi-static and frequency non-selective parameters. Consequently, the complex fading channel coefficient $g$ is a circular symmetric complex Gaussian random variable with zero mean and unit variance. Similarly, $\mathbf{f}$ and $\mathbf{h}$ are also uncorrelated circularly symmetric complex Gaussian random vectors, i.e., $\mathbf{f} = [f_1, \dots , f_M]^T  \sim  \mathcal {CN}  (0,\boldsymbol{I}_M$) and $\mathbf{h} = [h_1, \dots , h_M]^T \sim \mathcal{CN} (0,\boldsymbol{I}_M$). We make the following assumptions regarding the channels:
\begin{itemize}
    \item  The fading channel coefficients are assumed to be constant over the duration of one set of backscatter and power transfer phases, i.e., $T_b + T_p$ seconds and independent and identically distributed from one $T_b + T_p$ slot to the next. The use of such channels is in line with the recent work in this research field~\cite{Krikidis-2018, Lee-2018, Tao-2018}.
    \item We assume channel reciprocity, i.e., the channel from ER $\,\to\,$ ET during the backscatter phase and the channel from ET $\,\to\,$ ER during the power transfer phase are constant and transpose of each other~\cite{Lee-2018,gYang2014,xChen2014,hSon2014,park2015joint,jXu-2014,Zeng-2015}.
    \item In this work, we do not need to make any channel state information (CSI) assumption at the ET or the ER, as the retrodirective WPT technique precludes the need for CSI at either ET or the ER.
   \end{itemize}
\subsection{ Proposed Transmission Phases}
\indent The wireless power transfer from the ET to the ER takes place in two phases: (i) the backscatter phase and (ii) the power transfer phase, as shown in Fig.~\ref{smodel}. During the first backscatter phase of duration $T_b$, the ER initiates a request for WPT by sending a backscattered ambient signal to the ET. During the second power transfer phase of duration $T_p$, the ET performs retrodirective energy beamforming towards the ER. Note that in this work we will study the effect of varying the backscatter phase duration $T_b$, while we assume unit time in the power transfer phase.
\subsubsection{The Backscatter Phase}
\begin{figure}
\centering
\includegraphics[width=0.5  \textwidth]{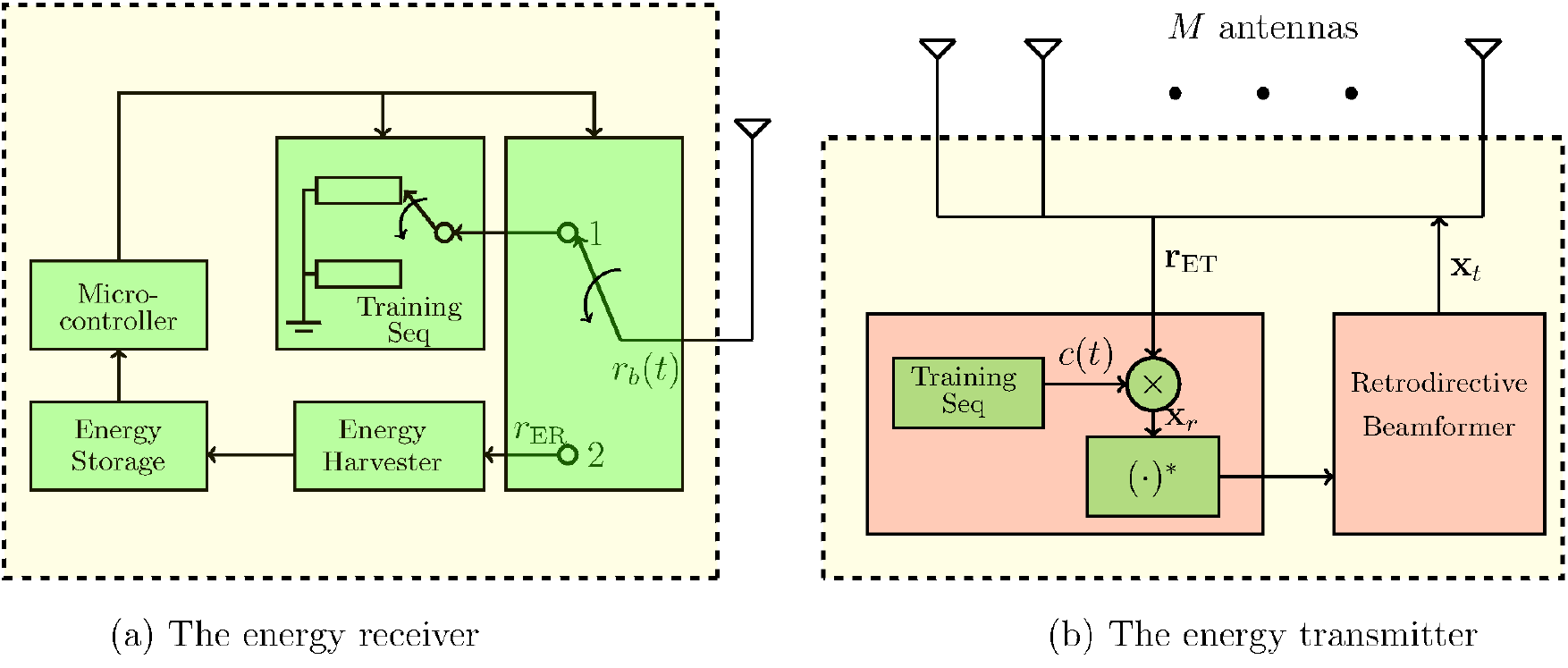}
        \caption{Block diagram of the energy transmitter and receiver.}
        \label{bdiag}
\end{figure}

\indent The backscattering  at the ER is achieved by adapting the level of the antenna impedance mismatch, which affects the power of the reflected signal. During the backscatter phase of duration $T_b$ seconds, the switch in Fig.~\ref{bdiag}a stays in position 1 and the ER backscatters the ambient signal given by $r_b(t) = \sqrt{\gamma_1} g \beta s(t)$ where $\beta$ is the backscatter reflection coefficient and $\sqrt{\gamma_1}gs(t)$ is the ambient signal arriving at the ER to be backscattered after suffering large scale attenuation $\gamma_1$ and channel coefficient $g$. In this work, we consider a BPSK-like backscatter coefficient having two different values, i.e., $\beta = \pm 1$.\footnote{$\beta$ can assume any pair of values $|\beta| \leq 1$. However, for simplicity we assume that $|\beta| = 1$.} The backscatter training means that the tag backscatters the ambient signal while switching the backscatter coefficient $N_c$ times\footnote{\sugcom{In practice, the switching of the backscatter coefficient would be activated using an oscillator. The state-of-the art low power backscatter tags have internal oscillators that consume only tens of microwatts of power~\cite{zhang2016enabling} and are feasible to be employed in our system model.}} according to a pre-defined sequence between the values $+1$ and $-1$ at a rate of $\frac{1}{T_c}$, where $T_c$ is the duration for which the backscatter coefficient maintains a certain value. This is effectively equivalent to multiplying the backscattered signal with a training signal $c(t)$ of $N_c$ short duration pulses of amplitude $+1$ and $-1$. Thus, at a given time instant $t$, the backscattered signal from the ER is given by $r_b(t) = \sqrt{\gamma_1} c(t) s(t)$, where $\gamma_1$, $g$ and $s(t)$ are as given above and $c(t)$ is the training signal composed of a sequence of $+1$ and $-1$ pulses governed by the backscatter coefficient. This training sequence applied at the ER is quite similar to the Direct Sequence Spread Spectrum (DSSS)~\cite{goldsmith-2005}\footnote{\sugcom{The signal backscattered from the ER is spread in frequency. However, its in-band and out-of-band interference to the licensed users is negligible since it is very weak, i.e., it is being generated by ambient backscatter and not active transmission~\cite{Huynh-2018}.}}. Henceforth, we will also refer to the short duration pulses of switching the reflection coefficient as `chips' and $T_c$ as the chip duration due to the similarity of this scenario with DSSS.\\
\indent The ET receives the composite signal consisting of the backscattered signal from the ER as well as the ambient signal and noise. The ET correlates this composite signal with the known training sequence $c(t)$. In this work, we assume perfect timing synchronization at the ET, in the baseline case. We then investigate the impact of imperfect synchronization in Section~\ref{imperf}.\\
\indent The purpose of using backscatter training is as follows. In general, the ambient signal is much stronger than the backscattered signal. This is because the latter suffers pathloss and attenuation twice and is orders of magnitude smaller than the former. The training performed at the ER before backscattering opens up a possibility for dealing with this issue of direct-link interference from the ambient signal at the ET. This is discussed in Section~\ref{detseq}.
\subsubsection{The Power Transfer Phase}
During the power transfer phase, the ET provides retrodirective wireless power transfer to the ER. Specifically, the ET conjugates the phase of the de-spread signal and each antenna at the ET sends a single-tone sinusoidal waveform towards the ER as shown in Fig.~\ref{bdiag}b. The phase and amplitude of this waveform are set according to the conjugated signal, subject to the maximum total transmit power $P_t$ at the ET. The switch in the ER in Fig.~\ref{bdiag}a moves to position 2. Consequently, the ER stops backscattering and only harvests energy from the energy beam directed to it by the ET. This energy is stored in the energy storage device in the ER. Note that during the backscatter phase when the ER is backscattering the ambient signals, the energy harvester remains idle and can complete the rectification and storage of energy.
\section{Signal Model}\label{sig_model}
In this section, we present the signal equations that form the basis of analysis and design in the later sections. We adopt a continuous-time baseband signal model.
\subsection{The Ambient Signal}
For simplicity, similar to the previous works~\cite{Tao-2018}, we model the ambient signal as
\begin{align}\label{ambient}
\ s(t) = \sqrt{P_s}\sum_{i=1}^{\infty} s_i p_{s}(t-iT_s),
\end{align}
where $s_i \sim  \mathcal {CN}  (0,1)$ and $p_{s}(t)$ is a rectangular pulse of duration $T_s$ given by
\begin{align}\label{pulse_amb}
p_s(t) =\begin{cases}
             1,  & 0\leq t\leq T_s\\
             0,  & t > T_s.
      \end{cases}
\end{align}
Note that the power of an ambient symbol in~\eqref{ambient} is $P_s$.
\subsection{The Backscatter Phase}
In the backscatter phase, as described in Section II, the backscattered signal from the ER is given by
\begin{align}\label{back_sig}
\ r_b(t) = \sqrt{ \gamma_1} g c(t)s(t),
\end{align}
where $c(t)$ is the training sequence with length $N_c$ and chip duration $T_c$. It can be modelled as
\begin{align}\label{chip_seq}
\ c(t) = \sum_{n=0}^{N_c-1} c_n p_c(t-nT_c),
\end{align}
where $c_n$ is the $n$-th chip ($+1$ or $-1$) of the training sequence and $p_c(t)$ is a rectangular pulse of duration $T_c$, i.e.,
\begin{align}\label{pulse_chip}
p_c(t) =\begin{cases}
             1,  & 0\leq t\leq T_c\\
             0,  & t > T_c.
      \end{cases}
\end{align}
The received signal at the ET is given by
\begin{align}\label{rec_sig}
 \mathbf{r}_\textrm{ET}(t) &= \sqrt{\gamma_2}\mathbf{f} r_b(t)+ \sqrt{\gamma_3}\mathbf{h} s(t) + \mathbf{n}(t)\nonumber\\
&= \sqrt{\gamma_1 \gamma_2} g \mathbf{f} c(t)s(t)  + \sqrt{\gamma_3} \mathbf{h} s(t) + \mathbf{n}(t) ,
\end{align}
where $\mathbf{n}(t)\sim \mathcal{CN} (0,{\sigma_n}^2 \boldsymbol{I}_M)$ is the AWGN. Note that $\mathbf{r}_{ET}(t)$ is a composite signal with three components, i.e., the backscattered signal from the ER, the ambient signal from the AS and the AWGN.
The ET correlates this composite signal with the known training sequence with perfect frame synchronization to give
{\small \begin{align}\label{despread}
\mathbf{x}_{r} &= \frac{1}{\ N_cT_c} \int \limits_{0}^{N_cT_c} \mathbf{r}_\textrm{ET} (t) c(t) dt\nonumber\\
&= \underbrace{\frac{1}{\ N_cT_c}\int \limits_{0}^{N_cT_c} \sqrt{\gamma_1\gamma_2} g \mathbf{f} c(t)s(t)c(t)  dt}_{\mathbf{x}_{s}} \nonumber\\
&+ \underbrace{\frac{1}{\ N_cT_c}\int \limits_{0}^{N_cT_c} \sqrt{\gamma_3} \mathbf{h} s(t)c(t) dt}_{\mathbf{x}_{i}} + \underbrace{\frac{1}{\ N_cT_c}\int \limits_{0}^{N_cT_c} \mathbf{n}(t) c(t)dt}_ {\mathbf{\widetilde{n}}},
\end{align}} where $\mathbf{x}_{s}$ and $\mathbf{x}_{i}$ are desired signal and undesired ambient (i.e., interference) components at the output of the correlator. Substituting the value of $c(t)$ from~\eqref{chip_seq}, we get $\mathbf{x}_{s}$ and $\mathbf{x}_{i}$ as
{\small \begin{align}
\mathbf{x}_{s}&= \frac{\sqrt{\gamma_1 \gamma_2} g \mathbf{f}}{\ N_cT_c} \int \limits_{0}^{N_cT_c} s(t) \sum_{n=0}^{N_c-1} c_n p_c(t-nT_c)\sum_{m=0}^{N_c-1} c_m p_c(t-mT_c)
 dt,\nonumber\\
&=\frac{\sqrt{\gamma_1 \gamma_2} g \mathbf{f}}{\ N_cT_c} \int \limits_{0}^{N_cT_c}  \sum_{n=0}^{N_c-1}c_n^2 s(t){p_c}^2(t-nT_c)dt. \label{xs} \\
\mathbf{x}_{i} &= \frac{1}{\ N_cT_c} \int \limits_{0}^{N_cT_c} \sqrt{\gamma_3} s(t)c(t)\mathbf{h} dt, \nonumber\\
&= \frac{\sqrt{\gamma_3} \mathbf{h}}{ N_cT_c} \int \limits_{0}^{N_cT_c} s(t)\sum_{n=0}^{N_c-1} c_n p_c(t-nT_c)dt. \label{xi}
\end{align}}
\subsection{Power Transfer Phase}\label{pt_sig_model}
Once the received signal is correlated with the local copy of the training sequence, the phase of the signal at the output of the correlator in~\eqref{despread} is conjugated in accordance with the principle of retrodirective WPT. This conjugate signal then controls the phase and amplitude of ET's energy signal subject to the maximum total transmit power $P_t$ at the ET. It is given as in~\cite{Lee-2018},
\begin{align}\label{xt}
\mathbf{x}_t &= \sqrt{P_t} \frac{(\mathbf{x}_{r})^*}{\left\|\mathbf{x}_{r}\right\|},
\end{align}
where $\left\|\mathbf{x}_{r}\right\| = \sqrt{{\mathbf{x}_{r}}^T \mathbf{x}_{r}}$. Note that in~\eqref{xt}, we have dropped the time index $t$ because the baseband signal $x_t$ does not vary with time.
The signal received by the ER in the power transfer phase is given by
\begin{align}\label{rER}
{r}_{\textrm{ER}} &= \sqrt{\gamma_2}\mathbf{f}^T \mathbf{x}_t, \nonumber\\
&= \sqrt{\gamma_2 P_t} \frac{\left(\mathbf{f}^T  {\mathbf{x}_{s}}^*+ \mathbf{f}^T  {\mathbf{x}_{i}}^*+ \mathbf{f}^T {\mathbf{\widetilde{n}}}^*\right)}{\left\|\mathbf{x}_{s}+ \mathbf{x}_{i} + \mathbf{\widetilde{n}}\right\|},
\end{align}
where $\mathbf{x}_{s}$ is given in~\eqref{xs}, $\mathbf{x}_{i}$ is given in~\eqref{xi} and $ \mathbf{\widetilde{n}}\sim \mathcal{CN} (0,\frac{{\sigma_n}^2}{N_cT_c}\boldsymbol{I}_M)$ is the noise at the output of the matched filter. Note that the receiver noise at the ER is not included in~\eqref{rER} because it is irrelevant to energy harvesting.
\subsection{Non-linear Energy Harvester}
In this work, we have assumed that the ER is equipped with a non-linear energy harvester modelled as follows~\cite{Alevizos2018nonlinear,Boshkovska-2017,Clerckx-2019WIPT}. Assuming that the incident RF power on the ER is $Q_{RF} = |{r}_{\textrm{ER}}|^2$, where ${r}_{\textrm{ER}}$ is the received signal at the ER during power transfer phase as given in~\eqref{rER}, the instantaneous harvested power by the energy harvester in the ER is given by
\begin {align}\label{qi_def}
Q = \frac{\frac{c_0}{1+\exp(-a_0(Q_{RF}-b_0))}-\frac{c_0}{1+\exp(a_0b_0)}}{1 - \frac{1}{1+\exp(a_0b_0)}},
\end{align}
where the parameters $a_0$, $b_0$ and $c_0$ respectively reflect the nonlinear charging rate with respect to the input power, the minimum turn-on voltage and the maximal harvested power when the energy harvester is drawn into saturation.
\subsection{Metric}
In this work, we use the average harvested power at the ER, $\overline{Q}$, as the figure of merit. \sugcom{It is defined as
\begin {align}\label{q_def}
\overline{Q} = E[Q],
\end{align}
where $Q$ is the instantaneous harvested power given by~\eqref{qi_def}}.\footnote{In this work, we assume unit time in the power transfer phase. Hence, we use the terms energy and power interchangeably}.
\section{Analysis of Energy Harvested with a PN Sequence}\label{PNseq}
In this section, we discuss the ambient backscatter training performed at the ER. As explained before, the ET receives a backscattered ambient signal from the ER. In addition to this, the ET also receives the original ambient signal which is orders of magnitude stronger than its backscattered version from the ER. This is due to the fact that the backscatter signal suffers attenuation twice, i.e., in going from AS to ER and then from ER to ET. As a result, it is considerably weakened and the signal received at the ET during the backscatter phase is predominantly composed of the ambient component.\\
\indent This problem of recovering the weak backscatter signal in the presence of a much stronger unwanted ambient signal is quite similar to the signal recovery problem in the direct sequence spread spectrum (DSSS). Taking inspiration from that, we consider a pseudo-noise (PN) training sequence at the ER when backscattering, i.e., the backscatter coefficient is switched between $+1$ and $-1$ in a pseudo-random fashion. By doing this, we expect to capitalize on the spreading gain and boost the backscatter signal against the direct-link ambient interference. In order to assess this technique and the impact of the spreading gain, we evaluate the power harvested at the ER during the power transfer phase of this scheme. We assume that the number of ambient symbols in the backscatter phase is $N_s$, i.e., $T_b = N_sT_s = N_cT_c$.\\
\indent We analyze the expressions for the desired signal component and the undesired ambient component to find the energy harvested by the ER in the following two cases: (i) $N_s \leq N_c$ and (ii) $N_s \geq N_c$. The main result is presented in the proposition below.\\
\begin{proposition}
For the system model considered in Section~\ref{sys_model} with Nakagami-$m$ fading channels when the number of antennas at the ET $M\,\to\, \infty$, the incident RF power on the ER is given by~\eqref{q_instant}
where
{\begin{align}
\mu &= \left|\sum_{i=1}^{N_s} s_i\right|^2 = \left|\sum_{i=1}^{N_s} s_i^*\right|^2,\label{mu}\\
\nu &= \left|\sum_{i=1}^{N_s} \sum_{n=\frac{N_c}{N_s}(i-1)}^{\frac{N_c}{N_s}i-1} c_n s_i^*\right|^2 = \left|\sum_{i=1}^{N_s} \sum_{n=\frac{N_c}{N_s}(i-1)}^{\frac{N_c}{N_s}i-1} c_n s_i\right|^2,\label{nu}
\end{align}}
for simplicity. Substituting this value of $Q_{RF}$ in~\eqref{qi_def}, we get the instantaneous harvested power at the ER, from which the average harvested power is calculated according to~(13).
\vspace{5mm}
\begin{figure*}[t]
\begin{equation}\label{q_instant}
 { Q_{RF} =
    \begin{cases}
      \gamma_2 P_t \left(\dfrac{\gamma_1 {\gamma_2}  |g|^2 \mu \left(M+\dfrac{1}{m_f}\right) +  \gamma_3 \nu \left(\dfrac{N_s}{N_c}\right)^2 + \dfrac{\sigma_n^2 N_s}{T_s P_s}} { \gamma_1 \gamma_2 |g|^2 \mu +  \gamma_3 \nu \left(\dfrac{N_s}{N_c}\right)^2  + \dfrac{\sigma_n^2 N_s}{T_s P_s}}\right) & \text{if $N_s \leq N_c$}\\
     \gamma_2 P_t \left(\dfrac{\gamma_1 {\gamma_2}  |g|^2 \mu \left(M+\dfrac{1}{m_f}\right) +  \gamma_3 \nu + \dfrac{\sigma_n^2 N_s}{T_s P_s}} { \gamma_1 \gamma_2 |g|^2 \mu + \gamma_3 \nu + \dfrac{\sigma_n^2 N_s}{T_s P_s}}\right)  & \text{if $N_s \geq N_c$}\\
    \end{cases}}
\end{equation}
\rule{18.2cm}{0.5pt}
\vspace{-5mm}
\end{figure*}
\end{proposition}
\begin{IEEEproof}
See Appendix~\ref{a}.
\end{IEEEproof}
\indent The general expression for the instantaneous harvested power in~\eqref{q_instant} has two mutually dependent random variables $\mu$ and $\nu$, in addition to $g$, $\mathbf{f}$ and $\mathbf{h}$. In addition, due to the nonlinear nature of the energy harvester, the overall expression for $Q$ in~\eqref{q_def} is fairly complex. Therefore, it is not possible to obtain a closed form expression for the expected value of harvested power. However, we can easily find the average harvested power by numerically taking the average of~\eqref{q_instant} substituted in~\eqref{qi_def} over a large number of Monte carlo realizations. Our simulation results in Section~\ref{res} confirm the accuracy of this approach.\\
\indent We have presented the average harvested power for the two possible cases of $N_s \leq N_c$ and $N_s \geq N_c$ in~\eqref{q_instant}. However, we will show in Fig.~\ref{eh1} in Section~\ref{res} that the harvested power becomes very low with increasing values of $N_s$. As $N_s$ exceeds $N_c$, the average harvested power stays perpetually low. This is due to the fact that the proposed scheme depends upon the variation of the backscatter coefficient during each ambient symbol that is backscattered. Therefore, from this point onwards, we only consider the case $N_s \leq N_c$.\\
\indent From the results in Fig.~\ref{eh1} in Section~\ref{pn_res}, the main conclusion is that even with the training sequence at work, the value of average harvested power is very small and it actually decreases with the increase of training duration. This is due to the fact that the ambient signal is orders of magnitude stronger than the backscattered signal. The spreading gain of the training sequence employed is not sufficient to boost the backscatter signal significantly against the ambient signal. Thus, during the power transfer phase, most of the energy transmitted by the ET effectively leaks towards the AS. Since the PN-sequence approach for training design fails to boost up the backscattered signal in the presence of the strong ambient interference, another approach of training sequence design is considered in the next section, that directly looks at eliminating the ambient interference. This new scheme relies on the variation of the backscatter coefficient between $\pm 1$ during each ambient symbol.
\section{The Proposed Training Sequence Design}\label{detseq}
As mentioned in the previous section, the purpose of employing backscatter training was to enable the ET to differentiate the backscattered transmission from the ambient signal. However, since the backscattered signal is orders of magnitude weaker than the ambient interference and the DSSS approach cannot boost up the backscatter signal, the only option left is to directly cancel or significantly suppress the ambient interference. In the following, we propose a scheme to remove the direct-link interference from the AS.\\
\indent \underline{\textit{Design Criterion:}} For the system model considered in Section~\ref{sys_model}, the ambient component can be eliminated at the output of the correlator in the ET if for each ambient symbol that is backscattered from the ER during the backscatter phase, the number of $+1$ and $-1$ chips is equal, i.e., $N_{+1} = N_{-1}$ and $N_{+1} + N_{-1} = \frac{N_c}{N_s}$, where $N_{+1}$ and $N_{-1}$ are the number of positive and negative chips respectively that are multiplied per symbol of the ambient source. This means that the backscatter coefficient is switched between $+1$ and $-1$ an even number of times, i.e., $N_c = 2kN_s$ where $k$ is a positive integer.\\
\indent We justify the above design criterion as follows:
In this case, $c(t)$ is a deterministic sequence of equal number of $+1$ and $-1$ chips instead of a PN sequence. Any sequence with equal number of $+1$ and $-1$ chips applied to each ambient symbol while backscattering, does the job. So we consider the expressions for $\mathbf{x}_\textrm{s}$ and $\mathbf{x}_\textrm{i}$, which are the expanded forms of~\eqref{xs} and~\eqref{xi} for $N_s \leq N_c$ (as derived in the Appendix), and are given below
\begin{align}\label{xsresolv}
\mathbf{x}_s &= \sqrt{\gamma_1 \gamma_2 P_s} \frac{ g \mathbf{f}}{N_s} \sum_{i=1}^{N_s} s_i.
\end{align}
\begin{align}\label{xiresolv}
\mathbf{x}_i &= \sqrt{\gamma_3 P_s} \frac{ \mathbf{h}}{N_c} \sum_{i=1}^{N_s} s_i \sum_{n=\frac{N_c}{N_s}(i-1)}^{\frac{N_c}{N_s}i - 1} c_n.
\end{align}
We can see from~\eqref{xsresolv} that, the desired backscattered component at the output of the correlator $\mathbf{x}_\textrm{s}$ does not depend on the attributes of the training sequence, i.e., how the backscatter coefficient is changed. Therefore, it remains the same as in the previous case. However, with our proposed training sequence satisfying the \textit{design criterion},~\eqref{xiresolv} becomes
\begin{align}
\mathbf{x}_i &= \sqrt{\gamma_3 P_s} \frac{ \mathbf{h}}{N_c} \sum_{i=1}^{N_s} s_i \sum_{n= \frac{N_c}{N_s}(i-1)}^{\frac{N_c}{N_s}i - 1} c_n, \nonumber\\
& =   \sqrt{\gamma_3 P_s} \frac{ \mathbf{h}}{N_c} \sum_{i=1}^{N_s} s_i \left[ (+1)N_{+1} + (-1)N_{-1} \right] =0
\end{align}
since $N_{+1} = N_{-1}$.
Thus, the ambient component at the output of the correlator cancels out.\\
\indent The following remarks discuss important practical aspects related to the \textit{design criterion}.
\begin{remark}\label{r4}
  The \textit{design criterion} is generic, i.e., any sequence that satisfies the two properties can serve the purpose. Moreover, we have seen that once the ambient component is removed, having a greater number of chips does not affect the harvested energy. Therefore, taking into account the hardware implementation, it is best to have the minimum number of chips per ambient symbol period, i.e., $k = 1$ and $N_c = 2N_s$ or $T_c = \frac{T_s}{2}$. This means that we can switch the backscatter coefficient only twice per ambient symbol, i.e., for each ambient symbol that is backscattered, the backscatter coefficient is kept $+1$ for half of the ambient symbol duration and $-1$ for the other half.
\end{remark}
\begin{remark}\label{r5}
It is interesting to see how this design criterion compares with the well-known training sequences commonly used in wireless communications, i.e., Maximal length sequences, Gold sequences, Walsh-Hadamard sequences and Kasami sequences. Out of these, only the Walsh-Hadamard sequences have equal number of $+1$ and $-1$ and hence satisfy the \textit{design criterion}.
\end{remark}
Using the proposed sequence in the \textit{design criterion}, we find the average harvested energy at the ER, which is presented in the proposition below.
\begin{proposition}\label{p2}
For the system model considered in Section~\ref{sys_model} with Nakagami-$m$ fading channels and $N_s \leq N_c$ while employing the backscatter training scheme proposed in the \textit{design criterion}, when the number of antennas at the ET $M\,\to\, \infty$, the incident RF power on the ER is given by
\begin{align}\label{q_instp}
Q_{RF} \approx \gamma_2 P_t \left(\frac{\gamma_1 \gamma_2 |g|^2 \mu \left(M+\dfrac{1}{m_f}\right) + \frac{\sigma_n^2 N_s}{T_s P_s}} {\gamma_1 \gamma_2 |g|^2 \mu  + \frac{\sigma_n^2 N_s}{T_s P_s}}\right).
\end{align}
where $\mu$ is as defined in~\eqref{mu}. Substituting this value of $Q_{RF}$ in~\eqref{qi_def}, we get the instantaneous harvested power at the ER, from which the average harvested power is calculated according to~\eqref{q_def}.
\end{proposition}
\begin{IEEEproof}
The proof is similar to the procedure in Appendix A and is omitted for the sake of brevity.
\end{IEEEproof}

\indent The following insight is gained from \textit{Proposition 2.}
\begin{remark}\label{r3}
  As the proposed scheme completely removes the direct-link ambient interference, the term involving the random variable $\nu$ is removed from the expression of $Q_{RF}$. Thus, during the power transfer phase, the ET forms a focussed beam towards the ER with no energy leaking towards the AS. This leads to a significant improvement in the harvested energy. This is demonstrated in the numerical results in Section~\ref{res}.
\end{remark}
\section{Impact of Practical System Imperfections}\label{imperf}
In the previous section, we propose a training design, under the perfect synchronization assumption. However, in practice, if the ambient symbol duration is unknown or changes from the one for which the system is designed, it may lead to the loss of timing synchronization at the correlator in the ET or unequal durations of $+1$ and $-1$ values of the backscatter coefficient at the ER. Consequently, the ambient signal may not be completely cancelled and the performance of the system in terms of average harvested power at the ER may be affected. In this section, we study the impact of the following practical system imperfections caused by the unknown duration of the ambient symbol.
\subsection{Imperfect synchronization at the correlator}\label{imperfsynch}
The analysis in Section~\ref{detseq} assumes perfect synchronization. In this sub-section, we consider the case when an integer number of ambient symbols fit in the duration of the  backscatter phase, but there is a misalignment between the received signal at ET and the locally generated training sequence during the backscatter phase. We model this misalignment as a time offset $T_{\textrm{off}}$.
\subsubsection{Effect of offset on the desired signal component}
\begin{figure}
\centering
\includegraphics[width=0.5  \textwidth]{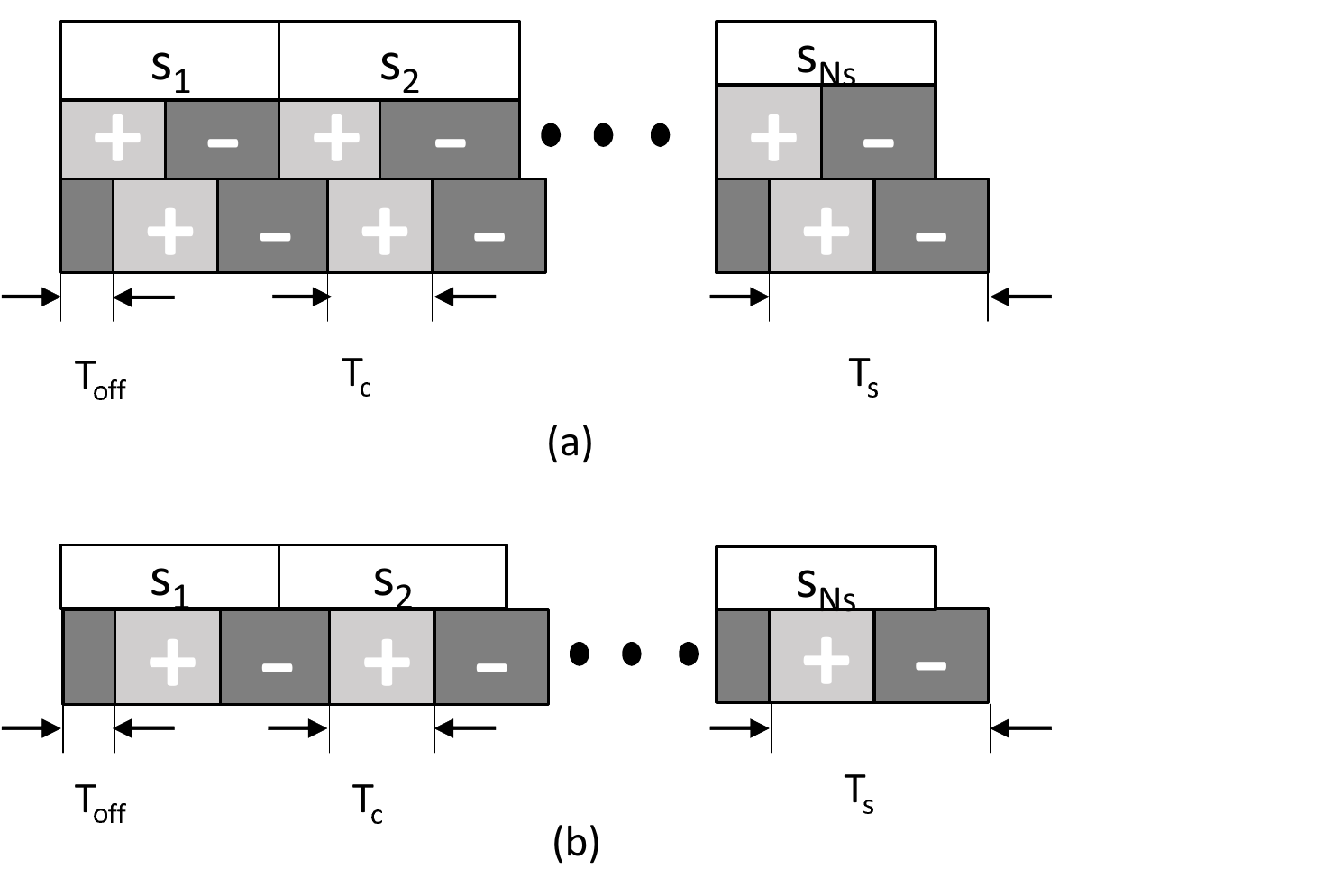}
        \caption{Misalignment between the backscattered signal and locally generated spreading sequence at the ET: (a) Effect on the backscatter component (b) Effect on the ambient component}
        \label{misal}
\end{figure}

We assume that the timing offset $T_{\textrm{off}} < T_c$. This is shown in Fig.~\ref{misal}.
In this case, the desired component at the output of the correlator in~\eqref{xs} becomes
\begin{align}\label{misxs}
\mathbf{x}_{s}&= \frac{\sqrt{\gamma_2 \gamma_1} g \mathbf{f}}{\ N_cT_c} \int_{0}^{N_cT_c}  \sum_{n=0}^{N_c-1}s(t)c_n p_c(t-nT_c)\nonumber\\
& \sum_{m=0}^{N_c-1}c_m p_c(t-T_{\textrm{off}}-mT_c) dt,\nonumber\\
& \displaystyle_{=}^{(a)} \sqrt{\gamma_1 \gamma_2 P_s} \frac{g \mathbf{f}}{N_cT_c} \sum_{i=1}^{N_s} s_i \frac{N_c}{N_s}\left( \int_{0}^{T_\textrm{off}}-1 dt + \int_{T_\textrm{off}}^{T_c} 1dt \right), \nonumber\\
& =  \sqrt{\gamma_1 \gamma_2 P_s} \frac{g \mathbf{f}}{N_cT_c} \sum_{i=1}^{N_s}s_i  \frac{N_c}{N_s} \left( -2 T_\textrm{off} + T_c \right),\nonumber\\
& =  \sqrt{\gamma_1 \gamma_2 P_s}\frac{g \mathbf{f}}{N_s}\sum_{i=1}^{N_s}s_i \left(1 - 2\frac{T_\textrm{off}}{T_c} \right),
\end{align}
\noindent where (a) splits the overall integration into intervals over each symbol.\\
\indent Comparing~\eqref{xs} and~\eqref{misxs} above we get for $T_\textrm{off} \leq T_c$
\begin{align}
\mathbf{x}_{s\textrm{(misaligned)}} = \left(1 - 2\frac{T_\textrm{off}}{T_c}  \right) \mathbf{x}_{s\textrm{(sychronized)}}.
\end{align}
Similarly, it can be shown that for $T_c < T_\textrm{off} \leq 2T_c$,
\begin{align}
\mathbf{x}_{s\textrm{(misaligned)}} = \left(2\frac{T_\textrm{off}}{T_c} -1  \right) \mathbf{x}_{s\textrm{(sychronized)}}.
\end{align}
Thus, we can see that if the synchronization is not perfect, the desired backscatter component is a fraction of the fully synchronized case.
\subsubsection{Effect of offset on the undesired ambient component}
Again assuming that the timing offset $T_{\textrm{off}} <T_c$, the undesired ambient component from~\eqref{xi} at the output of the correlator becomes,
{\small \begin{align}\label{misxi}
\mathbf{x}_{i}&= \frac{\sqrt{\gamma_3}  \textbf{h}}{\ N_cT_c} \int \limits_{0}^{N_cT_c} \sum_{m=0}^{N_c-1}c_m p_c(t-T_{\textrm{off}}-nT_c) s(t)dt,\nonumber\\
&=\frac{\sqrt{\gamma_3 P_s}  \textbf{h}}{\ N_cT_c} \int \limits_{0}^{N_cT_c} \sum_{m=0}^{N_c-1}c_m p_c(t-T_{\textrm{off}}-nT_c) \sum_{i=1}^{S} s_i p_{s}(t-iT_s)dt,\nonumber\\
& = \sqrt{\gamma_3 P_s} \frac{\textbf{h}}{N_cT_c} \sum_{i=1}^{N_s} s_i\nonumber\\
& \left(\int \limits_{0}^{T_\textrm{off}}-1 dt + \int \limits_{T_\textrm{off}}^{T_c} +1 dt + \int \limits_{T_c}^{2T_c} -1 dt + \cdots + \int \limits_{(\frac{N_c}{N_s}-1) T_c}^{\frac{N_c}{N_s} T_c -T_\textrm{off}} -1 dt \right),\nonumber\\
& = \sqrt{\gamma_3 P_s} \frac{\textbf{h}}{N_cT_c} \sum_{i=1}^{N_s} s_i \left( -T_\textrm{off} +T_c -T_c + T_c \cdots -T_c + T_\textrm{off} \right),\nonumber\\
& = 0.
\end{align}}
Note that the same result is obtained even when $T_{\textrm{off}} > T_c$.\\
\indent As we have seen in the previous sub-section, the desired component is scaled down because of the offset in synchronization while the undesired component is still completely being eliminated. This change in the magnitude of the desired component is reflected in the energy harvested at the ER. \textit{Therefore, we can conclude that the system can work reasonably well with a small timing offset. However good synchronization is needed for best performance.}\\
\indent Using the above values of $\mathbf{x}_s$ and $\mathbf{x}_i$, the incident RF power at the ER in case of misalignment at the ET can be shown to be given by
\begin{align}\label{q_instp}
Q_{RF} \approx \gamma_2 P_t \left( \frac{\left|1 - \frac{2T_{\textrm{off}}}{T_c}\right|^2\gamma_1 \gamma_2 |g|^2 \mu \left(M+\dfrac{1}{m_f}\right) + \frac{\sigma_n^2 N_s}{T_s P_s}} {\gamma_1 \gamma_2 |g|^2 \mu  + \frac{\sigma_n^2 N_s}{T_s P_s}}\right).
\end{align}
which holds for all values of $T_\textrm{off}$ except when $T_\textrm{off} = k\frac{T_c}{2}$, where $k$ is an integer and $k \geq 0$. Substituting this value of $Q_{RF}$ in~\eqref{qi_def}, we get the instantaneous harvested power at the ER, from which the average harvested power is calculated according to~\eqref{q_def}.
\subsection{Effect of change in ambient symbol duration}
In this subsection, we consider the case where due to unknown ambient symbol duration, an even number of chips or backscatter coefficient changes do not fit in each ambient symbol. Consider the scenario in which the system is designed for an ambient symbol duration $T_s$. However, when the system is actually deployed, the available ambient symbol has a different duration, i.e., $T_s'$. In this situation, it is difficult to present any analytical results. Hence, we will investigate its impact using simulations in Section~\ref{ts_res}.
\subsection{Effect of other interference from neighbouring ambient sources}
In this subsection, we consider the impact of interference on our system from neighbouring ambient sources. In particular, the application of the chipping sequence at the ET increases the bandwidth of the backscatter signal. Therefore, the ambient signals in neighbouring frequencies can potentially cause interference to the system.\\
\indent The interference signal can be from a variety of sources and can even include the backscattered versions of these interference signals. Compared with the interference signals directly received at the ET, their backscattered versions have much weaker strength (by several orders of magnitude) when they reach the ET. Therefore, we only consider the directly received interference signals. In this work, we have assumed the original ambient signal to follow a normal distribution, since the ambient signal may come from a variety of sources and is usually random. Therefore, we assume that the aggregate interference signal from other ambient sources in the same environment follows a zero-mean circularly symmetric complex Gaussian distribution~\cite{Qian-2017,Tao-2018}, i.e., $\mathbf{u}_i(t)\sim \mathcal{CN} (0,{\sigma_{i}}^2 \boldsymbol{I}_M)$ where ${\sigma_i}^2$ is the received interference power.\\
\indent The expression in~\eqref{rec_sig} for the received signal at the ET thus becomes,
\begin{align}\label{rec_sig2}
 \mathbf{r}_\textrm{ET}(t) &= \sqrt{\gamma_1 \gamma_2} g \mathbf{f} c(t)s(t) + \sqrt{\gamma_3} \mathbf{h} s(t) + \mathbf{u}_i(t) + \mathbf{n}(t) ,
\end{align}
As the ET correlates this composite signal with the known training sequence we get,
{\small \begin{align}\label{despread2}
\mathbf{x}_{r} &= \underbrace{\frac{\sqrt{\gamma_2} \mathbf{f}}{\ N_cT_c}\int\displaylimits_{0}^{N_cT_c} \sqrt{\gamma_1} g c(t)s(t)c(t)dt}_{\mathbf{x}_{s}} + \underbrace{\frac{1}{\ N_cT_c}\int\displaylimits_{0}^{N_cT_c}\sqrt{\gamma_3} \mathbf{h} s(t)c(t) dt}_{\mathbf{x}_{i}}\nonumber\\
& + \underbrace{\frac{1}{\ N_cT_c}\int\displaylimits_{0}^{N_cT_c} \mathbf{u}_i(t) c(t)dt}_ {\mathbf{\widetilde{u_i}}} + \underbrace{\frac{1}{\ N_cT_c}\int\displaylimits_{0}^{N_cT_c} \mathbf{n}(t) c(t)dt}_ {\mathbf{\widetilde{n}}},
\end{align}}
where $\mathbf{x}_{s}$ and $\mathbf{x}_{i}$ are desired signal and undesired primary ambient component and $\mathbf{\widetilde{u_i}}$ is the interference component from the neighbouring ambient signals, with $ \mathbf{\widetilde{u_i}}\sim \mathcal{CN} (0,\frac{{\sigma_i}^2}{N_cT_c}\boldsymbol{I}_M)$ and $ \mathbf{\widetilde{n}}\sim \mathcal{CN} (0,\frac{{\sigma_n}^2}{N_cT_c}\boldsymbol{I}_M)$ is the noise at the output of the matched filter.\\
\indent Using~\eqref{despread2}, the signal received at the ER, previously given by~\eqref{rER} becomes,
\begin{align}
{r}_{\textrm{ER}} &= \sqrt{\gamma_2}\mathbf{f}^T \mathbf{x}_t = \sqrt{\gamma_2 P_t} \frac{\left(\mathbf{f}^T  {\mathbf{x}_{s}}^*+ \mathbf{f}^T  {\mathbf{x}_{i}}^*+ \mathbf{f}^T {\mathbf{\widetilde{u_i}}}^*+ \mathbf{f}^T {\mathbf{\widetilde{n}}}^*\right)}{\left\|\mathbf{x}_{s}+ \mathbf{x}_{i} + \mathbf{\widetilde{u_i}}+ \mathbf{\widetilde{n}}\right\|},\label{rERb}
\end{align}
where $\mathbf{x}_t = \sqrt{P_t} \frac{(\mathbf{x}_{r})^*}{\left\|\mathbf{x}_{r}\right\|}$. Thus, the incident RF power at the ER with interference present can be shown to be given by,
\begin{equation}\label{qa2}
\resizebox{0.5\textwidth}{!}{$Q_{RF} \approx \gamma_2 P_t \left(\frac{P_s\gamma_1 {\gamma_2}  |g|^2 \mu \left(M+\dfrac{1}{m_f}\right) + P_s \gamma_3 \nu \left(\frac{N_s}{N_c}\right)^2+ \frac{\sigma_i^2 N_s}{T_s} + \frac{\sigma_n^2 N_s}{T_s}} {P_s\gamma_1 {\gamma_2}  |g|^2 \mu+ P_s \gamma_3 \nu \left(\frac{N_s}{N_c}\right)^2+ \frac{\sigma_i^2 N_s}{T_s} + \frac{\sigma_n^2 N_s}{T_s}}\right)$}
\end{equation}
Substituting this value of $Q_{RF}$ in~\eqref{qi_def}, we get the instantaneous harvested power at the ER, from which the average harvested power is calculated according to~(13). Generally, larger interference power leads to a degradation in the average harvested power, because our training design and interference cancellation is only targeted at the interference from the primary ambient signal, not the secondary interference signals from neighbouring ambient sources. This is numerically investigated in Section VII-D.
\section{Results}\label{res}
In this section we present the numerical and simulation results. In order to model a practical ambient backscatter scenario, we set the distances as follows: $d_1 = 200$ m, $d_2 = 10$ m, $d_3 = 200$ m~\cite{Huynh-2018}. The values of the rest of system parameters are: $d_0 = 1$ m, $k_0 = 0.001$, $M = 500$, $P_t = 1$ W, $P_s = 1$ W, $\sigma_n^2 = 10^{-18}$, $T_s = 5~\mu$s , $T_c = 500$ ns. For the non-linear energy harvester, we set $a_0 = 1500$, $b_0 = 0.0022$ and $c_0 = 24$ mW~\cite{Boshkovska-2017}. The choice of $T_c = 500$ ns ensures that multipath delay spread is negligible~\cite{bharadia2015backfi}. As mentioned in Section~\ref{sys_model}, we have assumed Nakagami-$m$ fading on all channel links. However, we can see from~\eqref{q_instant} and~\eqref{q_instp} and that the final analytical result only depends upon $m_f$. Hence, for the sake of simplicity, we have considered $m_h = m_g = 1$ for the AS to ET and AS to ER links and $m_f = 1$ and $m_f = 10$ for the ER to ET link. \redcom{We initially ignore the impact of other interference from neighbouring ambient signals, setting $\sigma_i = 0$ in Sections~\ref{pn_res}-~\ref{ts_res}, and then investigate the impact of such interference in Section~\ref{intf_res}.}
\subsection{Energy Harvested with a PN Sequence}\label{pn_res}
\begin{figure}
\centering
\includegraphics[width=0.5  \textwidth]{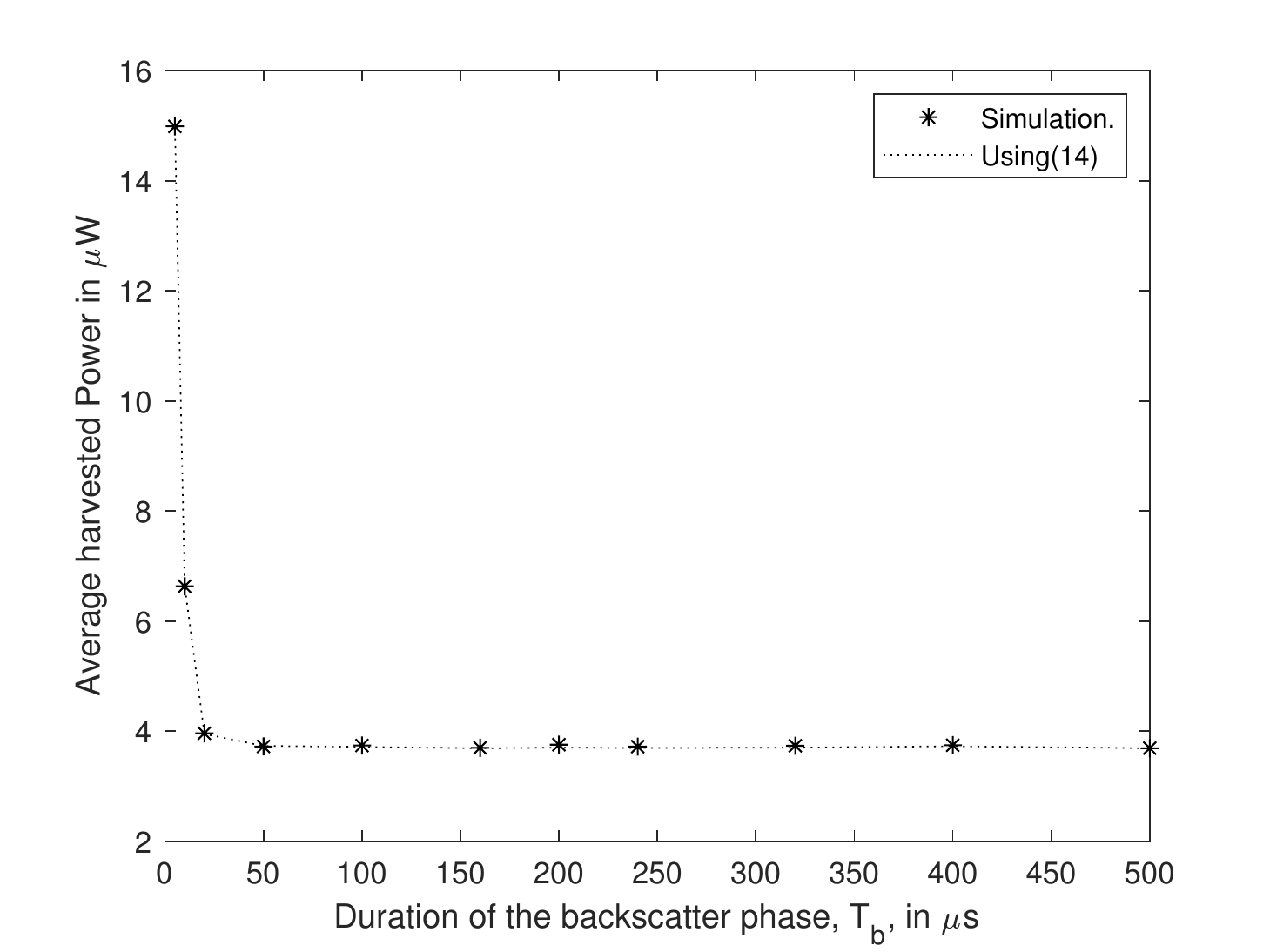}
        \caption{\sugcom{Average harvested power at the ER as a function of $T_b$ (duration of the backscatter phase).}}
        \label{eh1}
\end{figure}

\begin{figure}
\centering
\includegraphics[width=0.45  \textwidth]{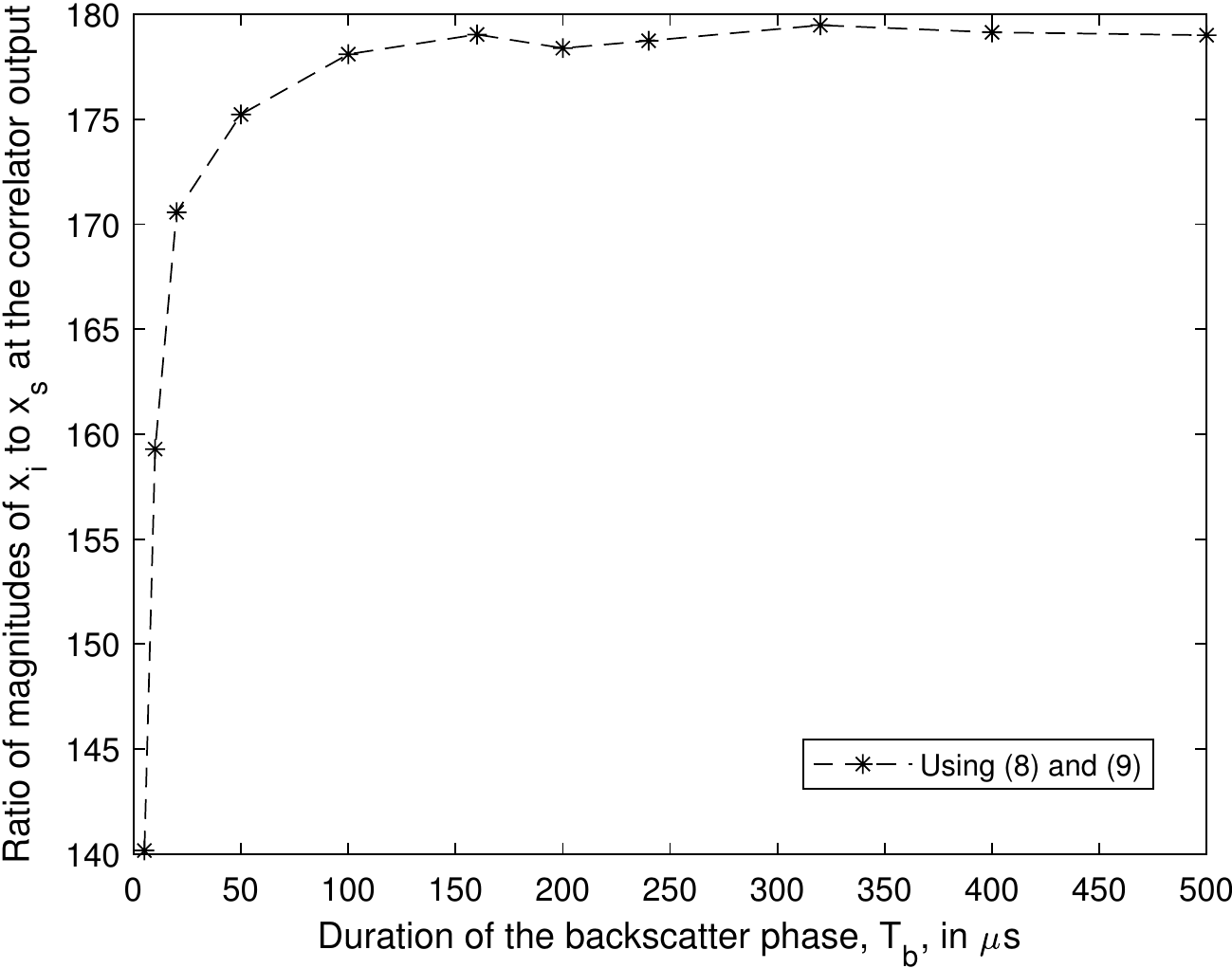}
        \caption{\sugcom{Ratio of magnitude of ambient and backscatter signal components at the output of the correlator plotted against $T_b$ (duration of the backscatter phase).}}
        \label{mgr}
\end{figure}

\indent Fig.~\ref{eh1} plots the average harvested power versus the duration of the backscatter phase, i.e., $T_b$ with the ambient signal duration being $T_s = 5~\mu$s. These results are averaged over $10^4$ Monte Carlo simulation trials. In each trial, a new pseudorandom sequence is generated and used. Note that for other practical values of system parameters, the average harvested power has very similar values and trend. Thus, we only show a single curve in Fig.~\ref{eh1}.\\
\indent The figure shows that there is a very good agreement between the analytical results in~\eqref{q_instant} and the simulation for $N_s \leq N_c$\footnote{\sugcom{A similar match is observed between the analytical result and the simulation for $N_s \geq N_c$ but the corresponding plots are not presented here due to the reason discussed in Section~\ref{PNseq}.}}. The figure also shows that the average harvested power is maximum around $15~\mu$W when $N_s = 1$ and $T_b = 5~\mu$s. As $N_s$ and hence $T_b$ increase, the average harvested power quickly decreases and reaches a value of approximately $4~\mu$W. Thus, we can conclude from Fig.~\ref{eh1} that the average harvested power is very small and it reduces further as the training period increases.\\
\indent  This  latter observation is particularly counter-intuitive, since it is not expected to happen when using DSSS techniques. The reason for this trend is that the ambient signal is orders of magnitude stronger than the backscattered signal. The spreading gain of the training sequence employed is not sufficient to boost the backscatter signal significantly against the ambient signal. In order to demonstrate this, Fig. ~\ref{mgr} plots $\frac{|x_i|}{|x_s|}$, i.e., the ratio of the magnitudes of the undesired ambient component and the desired backscatter component at the output of the correlator versus the duration of the backscatter phase $T_b$. We can see from the figure that even with the training sequence in use, the ambient component is much stronger than the desired backscattered signal. Moreover, as the duration of the training phase increases, the ambient component becomes increasingly stronger. Thus, when the ET performs retrodirective WPT by taking the conjugate of the composite signal at the output of the correlator, the comparative strength of the ambient component is far greater than the backscattered one for larger durations of backscatter phase. Thus, most of the energy transmitted by the ET is still effectively leaking towards the AS and this situation becomes exacerbated for longer durations of backscatter phase due to the comparatively higher strength of the ambient component.
\subsection{Energy Harvested with the Proposed Ambient Backscatter Training Scheme}\label{prop_res}
\begin{figure}
\centering
\includegraphics[width=0.5  \textwidth]{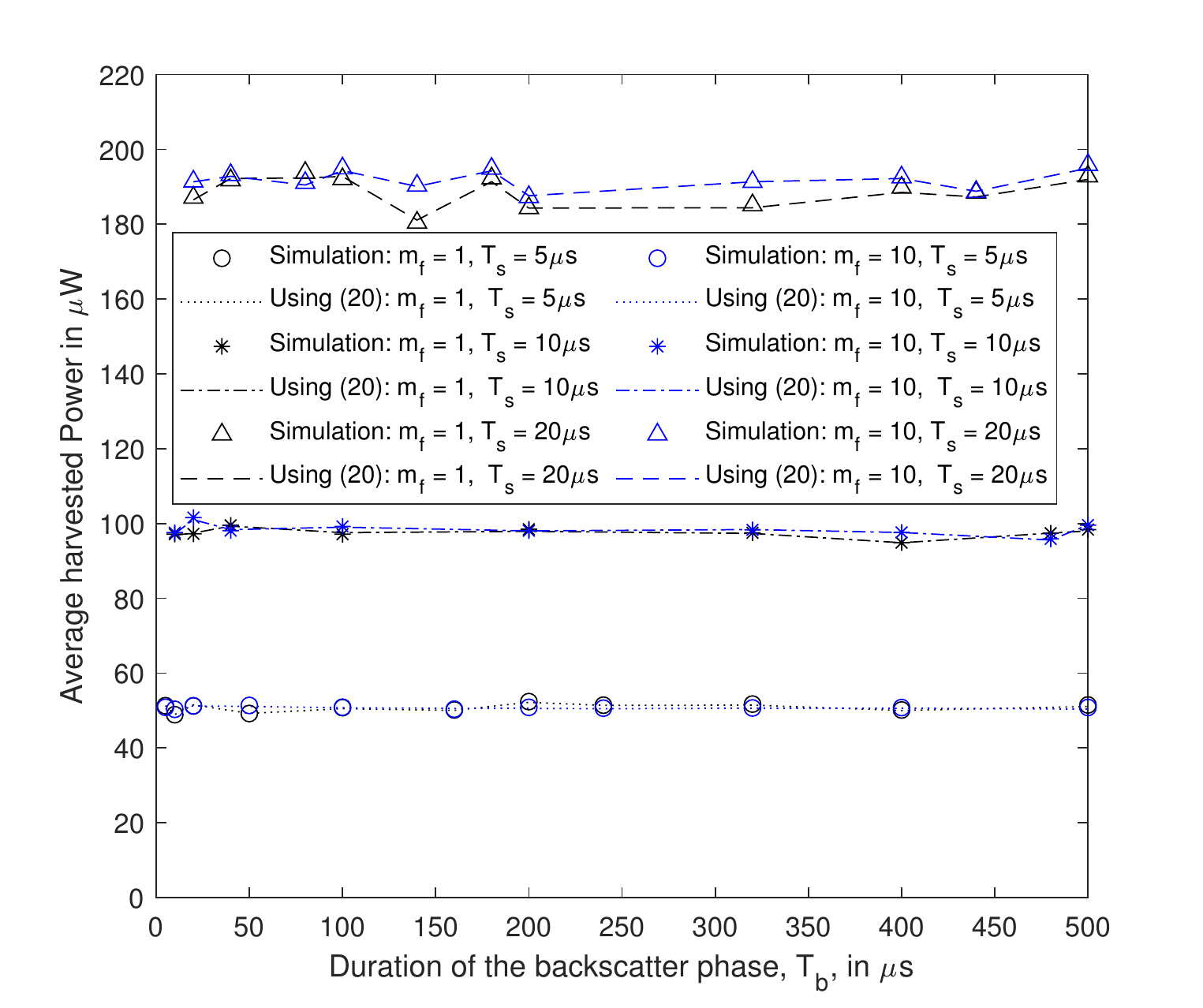}
        \caption{Average harvested power at the ER with the proposed sequence plotted against the duration of the backscatter phase, $T_b$.}
        \label{eh2}
\end{figure}

\begin{figure}
\centering
  \includegraphics[width=.5 \textwidth]{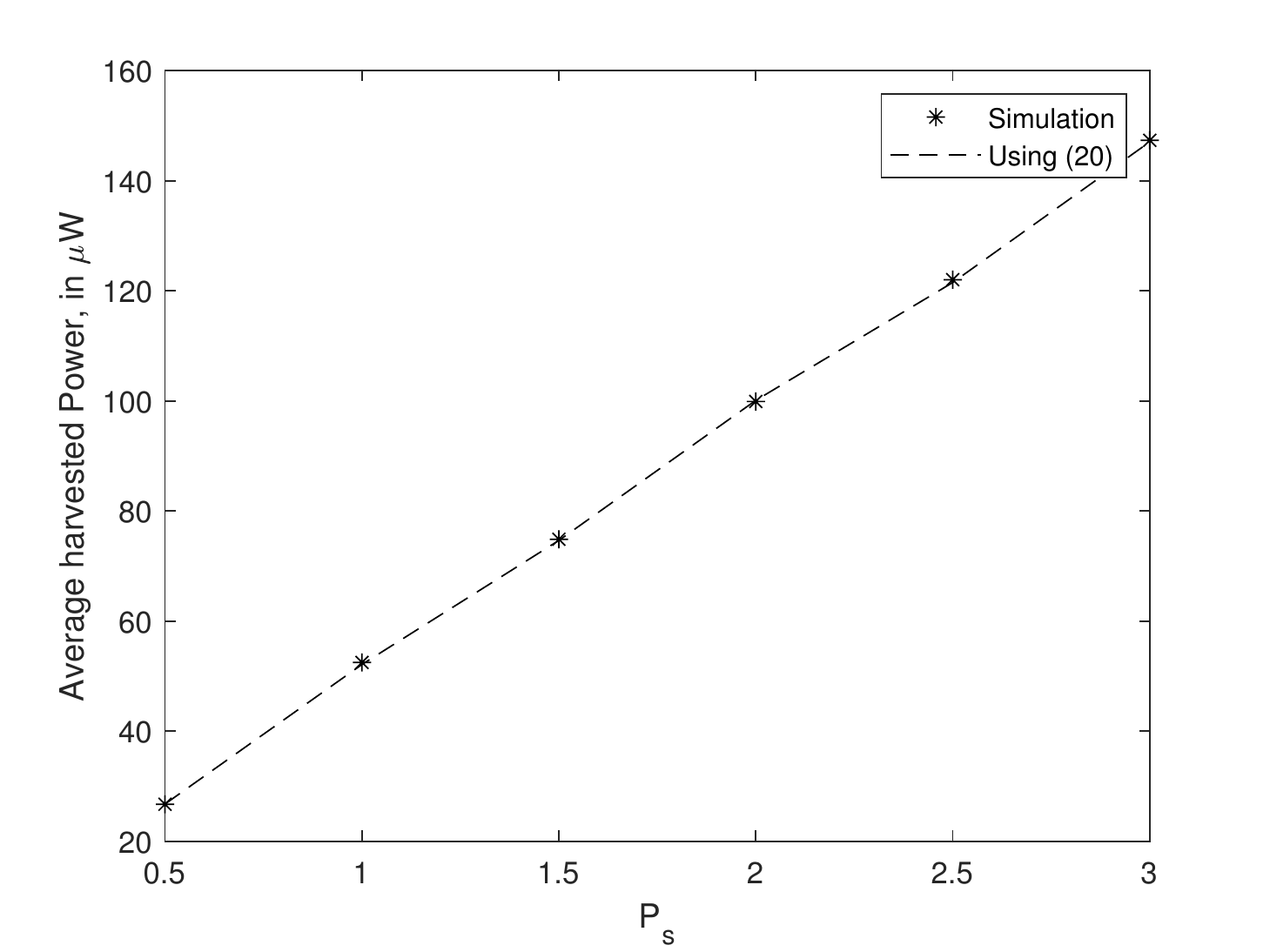}
  \caption{\sugcom{Average harvested power, $\bar{Q}$, \\ plotted against transmit power of AS, $P_s$.}}
  \label{ps_plots}
\end{figure}

\begin{figure}
  \centering
  \includegraphics[width=.5 \textwidth]{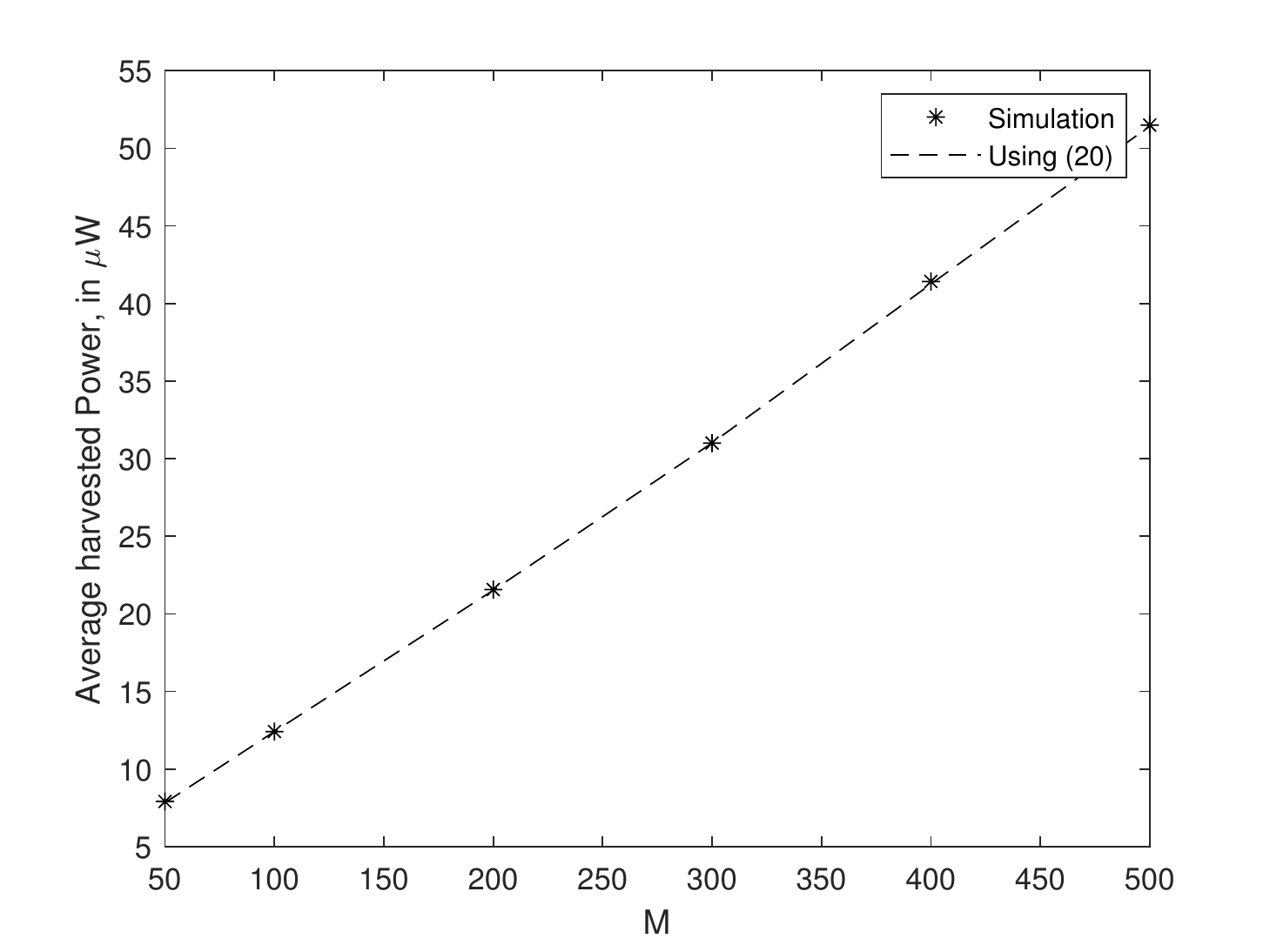}
  \caption{\sugcom{Average harvested power, $\bar{Q}$, \\ plotted against number of antennas at the ET $M$.}}
  \label{M_plots}
\end{figure}

\sugcom{Using~\eqref{q_instp},~\eqref{qi_def} and~\eqref{q_def}, the average harvested energy at the ER is calculated and plotted in Fig.~\ref{eh2} for three different values of ambient symbol duration, i.e., $T_s = 5~\mu$s, $T_s = 10~\mu$s and $T_s = 20~\mu$s and two different values of Nakagami-$m$ fading on the ET to ER link i.e. $m_f = 1$ and $m_f = 10$. The values of other system parameters are the same as stated in the beginning of this section for Fig.~\ref{eh1}. Numerous features of the proposed scheme are evident from Fig.~\ref{eh2}. Firstly, we can see that the result for $m_f = 1$ and $m_f = 10$ are quite similar. Thus, in this case, having a line of sight link between ET and ER does not significantly impact the results. Hence, in the remaining results, we only consider $m_f = 10$.}\\
\indent Secondly, it can be observed that the energy harvested at the ER increases significantly as compared to the case when a pseudo-random sequence is employed at the ER during the backscatter phase. This is due to the fact that the proposed scheme completely eliminates the ambient component. As a result, during retrodirective WPT the ET forms a focused beam directed back at the ER alone, with no energy leaking to the AS.\\
 \indent Thirdly, the harvested power at the ER does not change with the increase in backscatter training duration $T_b$, but stays constant as long as the ambient symbol duration $T_s$ stays constant. Specifically, when the system is designed with a fixed value of $T_c$, then for different values of $N_s$ and hence $T_b$, the average harvested power at the ER now stays around $50~\mu$W for $T_s = 5~\mu$s, $99~\mu$W for $T_s = 10~\mu$s and $190~\mu$W for $T_s = 20~\mu$s.\\
 \indent Fourthly, with the ambient component removed, the average harvested power depends largely on the duration of the ambient symbol $T_s$, as is evident from the plot with the average harvested power having a significantly larger value for $T_s = 20~\mu$s, compared to $T_s = 5~\mu$s \\
  \indent Lastly, it can also be inferred from the plot that for a fixed ambient source, the average harvested power in this case is independent of the number of chips $N_c$. Actually, for a fixed chip duration, the number of chips also increases with the increased backscatter period $T_b$ and as we can see from Fig.~\ref{eh2}, the average harvested power stays constant for the increased values of the backscatter period.\\
  \indent Fig.~\ref{ps_plots} presents the plots of average harvested power against the duration of backscatter phase $T_b$ for different values of $P_s$, the power of the AS. We can see that the average harvested power is larger with higher values of $P_s$. This observation is consistent with the analysis in Section~\ref{detseq}. We can see from~\eqref{xsresolv} that the desired backscatter component $\mathbf{x}_\textrm{s}$ is directly proportional to the strength of the AS. Since the ambient component is now completely removed, a higher value of power is harvested on average at the ER when the AS is stronger. Similarly, Fig.~\ref{M_plots} plots the average harvested power against $M$, the number of antennas at the ET. It can be observed that there is a good agreement again between the results obtained by simulation and by numerically averaging~\eqref{q_instp} for practical values of $M$.
\subsection{Impact of Practical System Imperfections}\label{imperf_res}
\subsubsection{Imperfect synchronization at the correlator}\label{synch_res}
\begin{figure}
\centering
\includegraphics[width=0.5  \textwidth]{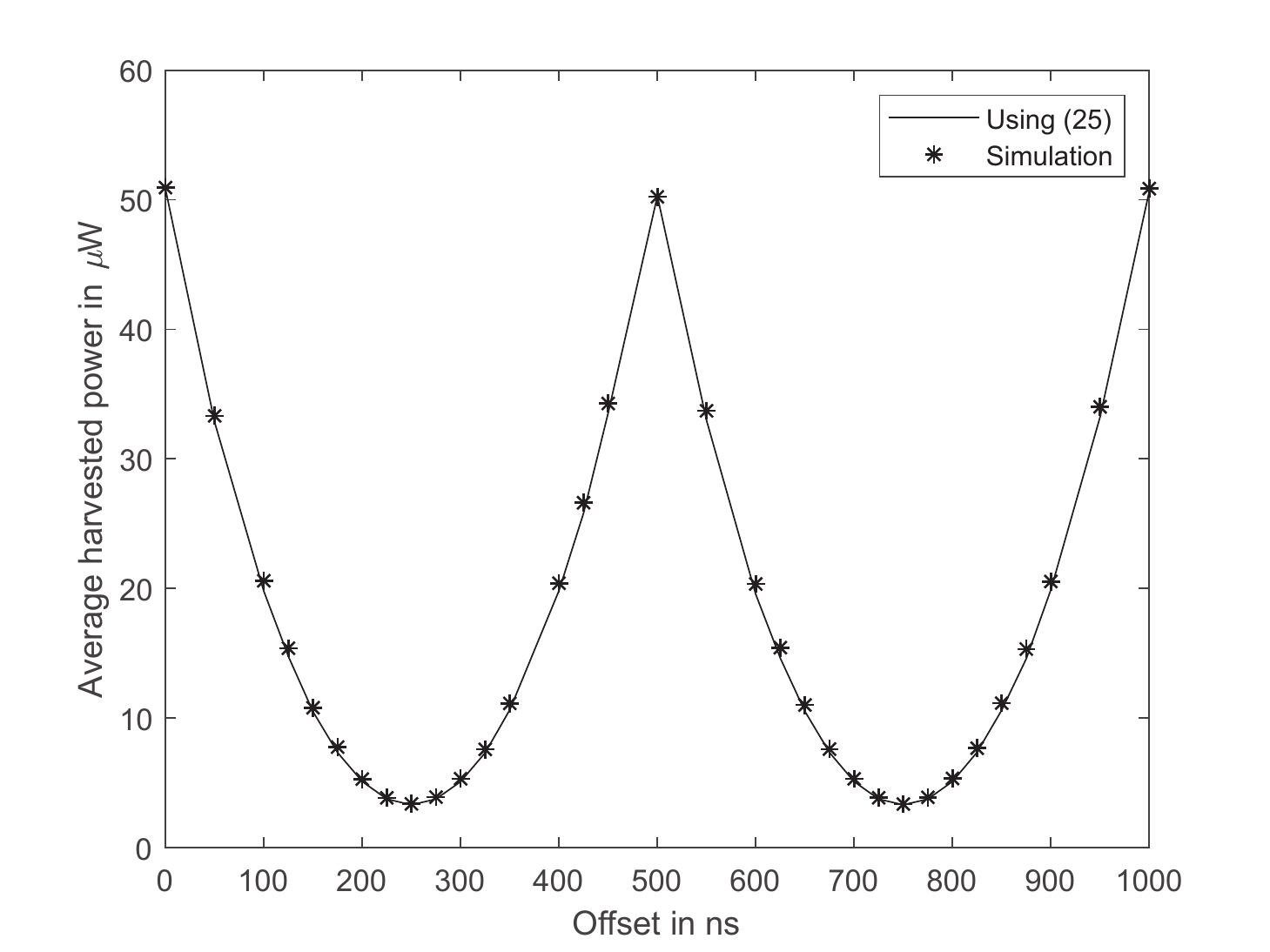}
        \caption{\sugcom{Average harvested power at the ER plotted against the offset between incoming and locally generated signal at the correlator.}}
        \label{misal2}
\end{figure}

A plot of the average harvested power at the ER as a function of time offset between the received and locally generated signal is given in Fig.~\ref{misal2}. The parameter values used are the same as for Fig.~\ref{eh2}. For this plot we have taken $T_c = \frac{T_s}{2}$, as discussed in Remark 1 in Section~\ref{detseq}. It can be seen from Fig.~\ref{misal2} that the average harvested power decreases for $kT_c \leq T_\textrm{off} < k\frac{T_c}{2}$, achieving a local minimum at $T_\textrm{off} = k \frac{T_c}{2}$ and then increases for $ k\frac{T_c}{2} < T_\textrm{off} \leq kT_c $. This reiterates that the system can work reasonably with a small offset, as discovered in Section~\ref{imperfsynch}.
\subsubsection{Effect of unknown ambient symbol duration}\label{ts_res}
\begin{figure}
\centering
\includegraphics[width=0.5  \textwidth]{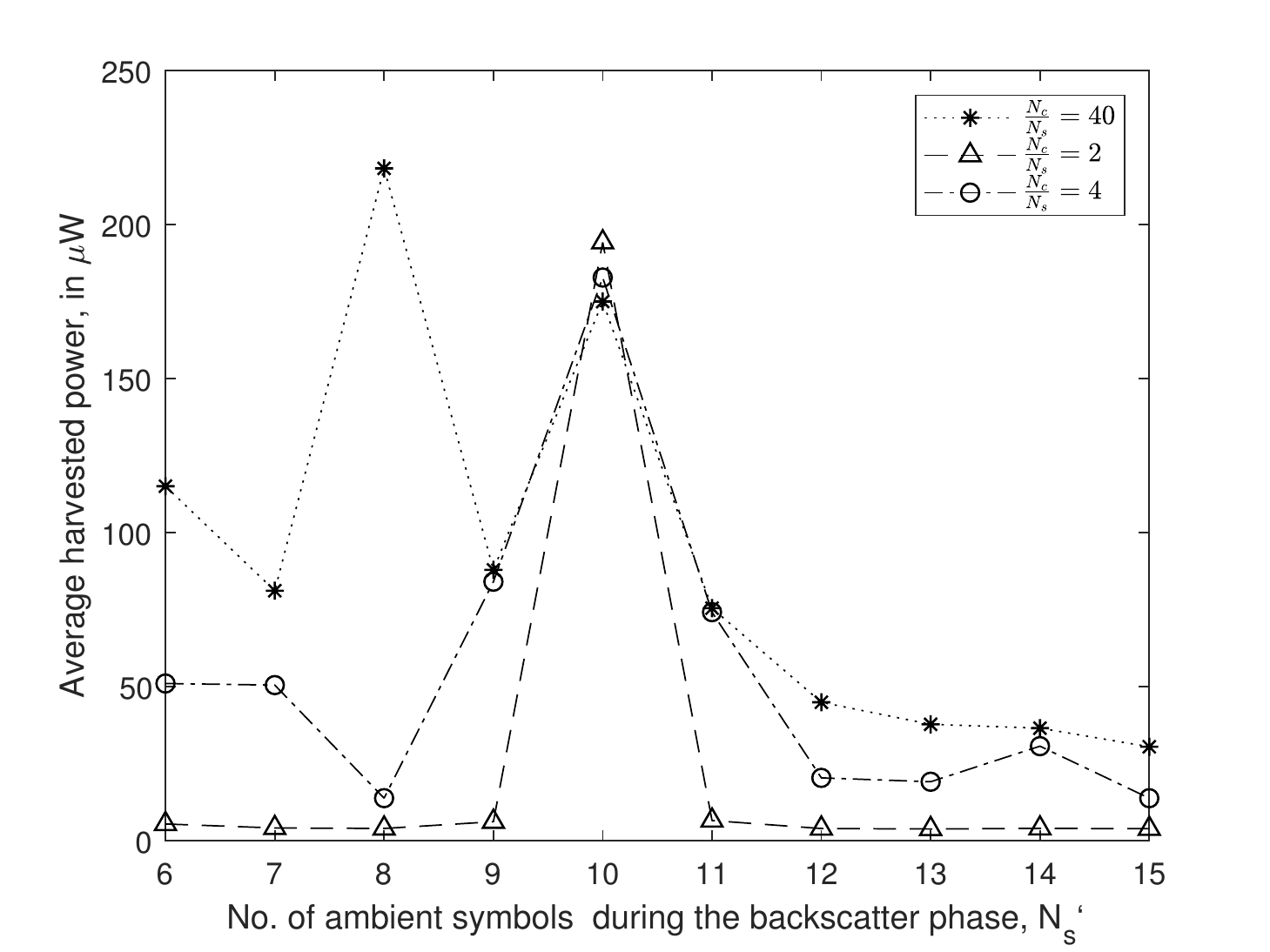}
        \caption{\sugcom{Average harvested power at the ER plotted against the number of ambient symbols during the backscatter phase when the actual ambient symbol duration is different from the designed value.}}
        \label{gensim}
\end{figure}

Fig.~\ref{gensim} plots the average harvested power at the ER versus the number of ambient symbols that fit in the backscatter phase duration of $T_b$ seconds for training sequences that satisfy the \textit{design criterion} but have different number of chips, i.e., $\frac{N_c}{N_s} = 2, 10$ and $40$. \sugcom{This system was originally designed for the following values: $T_b = 200~\mu$s, $N_c = 400$, $T_c = 500~$ns, $T_s = 5~\mu$s, $N_s = 10$, $m_g = m_h = 1$ and $m_f = 10$.} We plot the average harvested power at the ER for a range of values of $N_s' = \{6,7,8,9,10,11,12,13,14,15\}$ and the corresponding $T_s'$. \\
\indent We can see from Fig.~\ref{gensim} that the training sequence with the least number of chips per symbol gives the worst performance, i.e., as the number of the ambient symbols in the backscatter phase deviates from designed value, the average harvested power drops to a fraction of a $\mu$W. This is due to the fact that the ambient component is no longer completely cancelled as the ER was designed to switch the backscatter coefficient at $\frac{T_s}{2}$, so that each ambient symbol was multiplied by $+1$ and $-1$ for alternate halves of its duration. However, in the new scenario, a switch at $\frac{T_s'}{2}$ is required. Consequently, the ambient component is not eliminated completely; rather a fraction from each ambient symbol remains that contributes to a residual ambient component at the output of the correlator. This, in turn leads to a significant amount of power leaking to the AS. \\
\indent It can also be observed from Fig.~\ref{gensim} that as the number of chips per ambient symbol increase, better performance can be obtained. For instance, the curve with the largest number of chips per symbol, i.e., $\frac{N_c}{N_s} = 40$ performs relatively better than the other two cases for moderate mismatch in symbol duration. The reason for this behaviour is that the fraction of ambient component that is not cancelled due to the unknown value of $T_s$ depends upon the chip duration $T_c$. Therefore, in spite of the fact that an even number of chips may not fit in one ambient symbol (leading to imperfect cancellation), by increasing the switching rate of the backscatter coefficient and thereby decreasing the chip duration $T_c$, the un-cancelled fraction of a chip can be reduced and hence a smaller ambient component remains at the output of the correlator. In this way, there is less leakage towards the AS and the ER is able to harvest more power. \textit{Thus, when the ambient symbol duration is unknown, a faster switching rate can help to minimize the effect of uncancelled ambient for moderate mismatch of symbol duration.}\\
\subsection{Effect of other interference from neighbouring ambient sources}\label{intf_res}
\begin{figure}
\centering
\includegraphics[width=0.5  \textwidth]{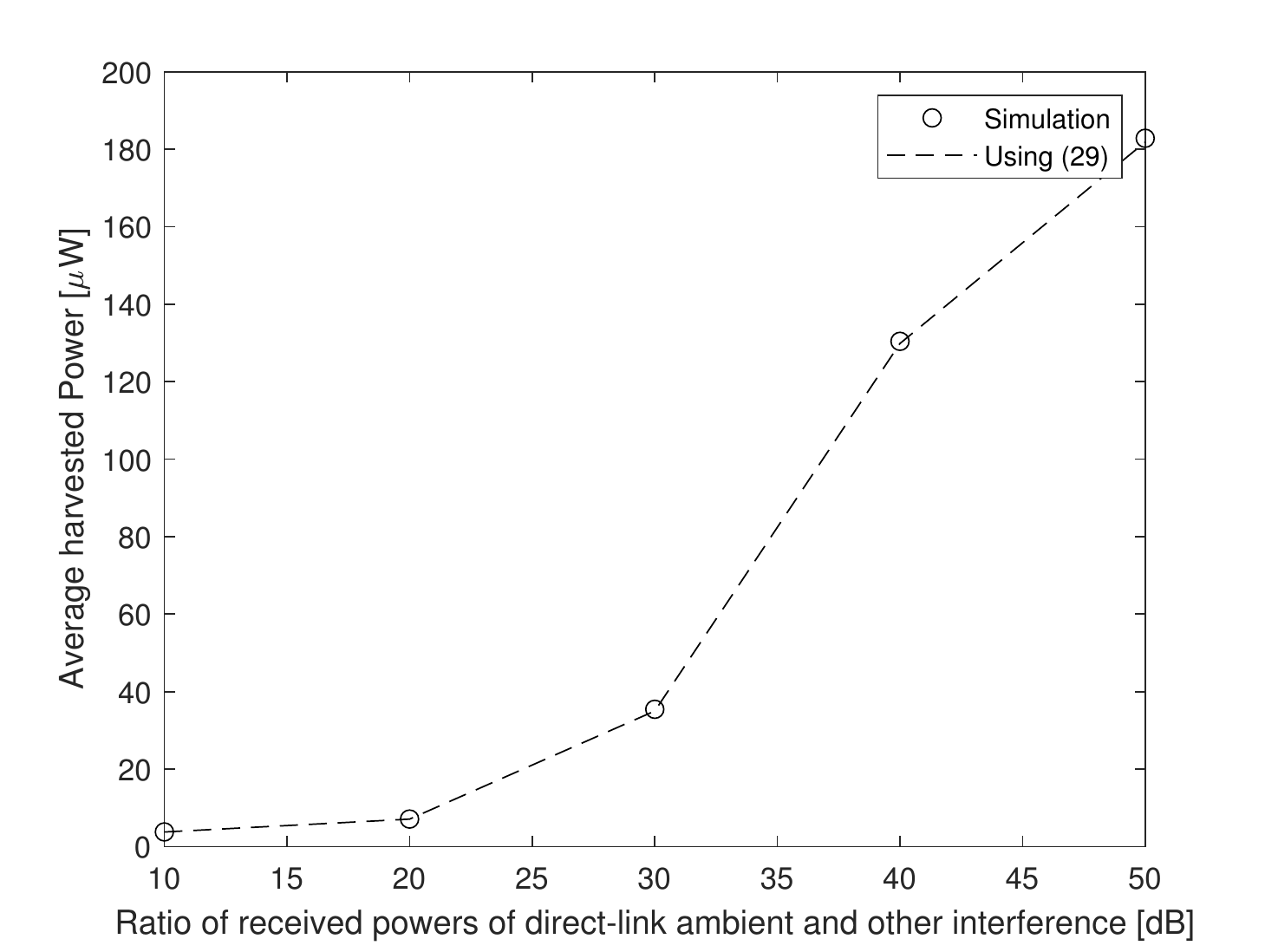}
        \caption{\redcom{Average harvested power at the ER versus the ratio of the average received power from the direct-link ambient and the average interference power from neighbouring ambient sources.}}
        \label{intf}
\end{figure}

\redcom{Fig.~\ref{intf} plots the average harvested power at the ER versus the ratio of the average received power from the direct-link AS and the received interference power from neighbouring sources $\sigma_i^2$. This ratio is expressed in dB. For this plot, we have taken $T_s = 20~\mu$s and $N_s = 4$ while all the other system parameters are kept the same as for Fig.~\ref{eh2}. It can be seen that the average harvested power is $7.09~\mu$W when this ratio is 20 dB. However, when this ratio increases to 30 dB and 40 dB, the average harvested power jumps to tens and hundreds of $\mu$W respectively, finally approaching the value of over $180~\mu$W for 50 dB, very close to that can be achieved when there is no interference. Therefore, if the interference signal is significantly weaker than the original ambient signal, our system can harvest tens to hundreds of $\mu$W of power.}
\section{Conclusions and Future Work}
In this work we have presented a wireless power transfer scheme to energize an ER using retrodirective WPT at the ET and ambient backscatter at the ER. To deal with the direct-link ambient interference, we have proposed the approach of backscatter training, i.e., the pattern of varying the reflection coefficient at the ER to completely eliminate the strong direct-link ambient interference. We have showed that when the ambient symbol duration is known, the switching rate does not matter and we can switch the backscatter coefficient only twice per ambient symbol period. When the ambient symbol duration is unknown, then switching at a faster rate helps to minimize the effect of the uncancelled ambient and boost the harvested power. \redcom{The best average harvested power is achieved when the interference signal from neighbouring ambient sources is significantly weaker than the original ambient signal.} The scheme proposed in this paper can be extended to multiple backscatter tags located in an area by assigning the mutually orthogonal Walsh-Hadamard sequences to individual ERs and considering scheduling or collision resolution schemes. This is outside the scope of this work and can be considered in future work.
\begin{appendices}
\section{Proof of Proposition 1}\label{a}
We derive the formula for instantaneous energy harvested at the ER during the power transfer phase as given in~\eqref{q_instant}. We consider the following  two cases:\\
\underline{Case 1: $N_s \leq N_c$}
In this case, we have $T_s \geq T_c$. Substituting~\eqref{ambient} in~\eqref{xs} we have
{\small \begin{subequations}
\begin{alignat}{4}
\mathbf{x}_{s} &= \frac{\sqrt{\gamma_1 \gamma_2 P_s}g \mathbf{f}}{N_cT_c} \int \limits_{0}^{N_cT_c} \sum_{n=0}^{N_c-1}c_n^2 {p_c}^2(t-nT_c) \sum_{i=1}^{N_s} s_i p_{s}(t-iT_s)dt, \label{xs1a}\\
&= \frac{\sqrt{\gamma_1 \gamma_2 P_s}g \mathbf{f}}{N_cT_c} \sum_{i=1}^{N_s} s_i \sum_{n=\frac{N_c}{N_s}(i-1)}^{\frac{N_c}{N_s}i-1} c_n^2 \int \limits_{nT_c}^{(n+1)T_c}  {p_c}^2(t-nT_c) dt,\label{xs1b}\\
&= \sqrt{\gamma_1 \gamma_2 P_s} \frac{ g \mathbf{f}}{\ N_cT_c } \sum_{i=1}^{N_s} \sum_{n=\frac{N_c}{N_s}(i-1)+1}^{\frac{N_c}{N_s}i} c_n^2 s_i T_c, \label{xs1cb}\\
&= \sqrt{\gamma_1 \gamma_2 P_s} \frac{ g \mathbf{f}}{\ N_c } \frac{N_c}{N_s} \sum_{i=1}^{N_s} s_i, \label{xs1d} \\
&= \sqrt{\gamma_1 \gamma_2 P_s} \frac{ g \mathbf{f}}{N_s} \sum_{i=1}^{N_s} s_i, \label{xs1ed}
\end{alignat}
\end{subequations}}
\noindent where the integration in~\eqref{xs1b} comes from the fact that the integration in~\eqref{xs1a} is being performed for the product of two aligned rectangular pulses $p_c(t)$ and $p_s(t)$ where $T_s \geq T_c$ and the duration of integration is $N_cT_c$. Also,~\eqref{xs1d} follows from the fact that $c_n^2 = 1$ and $\sum_{n=\frac{N_c}{N_s}(i-1)}^{\frac{N_c}{N_s}i-1} = \frac{N_c}{N_s}$ for any given $i$.\\
\indent Next, substituting~\eqref{ambient} in~\eqref{xi} we get
{\small \begin{subequations}
\begin{alignat}{3}
 \mathbf{x}_{i} &=  \frac{\sqrt{\gamma_3 P_s} \mathbf{h}}{ N_cT_c} \int \limits_{0}^{N_cT_c} \sum_{i=1}^{N_s} s_i p_{s}(t-iT_s) \sum_{n=1}^{N_c} c_n p_c(t-nT_c)dt, \label{xi1a}\\
 &= \sqrt{\gamma_3 P_s} \frac{ \mathbf{h}}{N_cT_c} \sum_{i=1}^{N_s}s_i \sum_{n=\frac{N_c}{N_s}(i-1)}^{\frac{N_c}{N_s}i-1} c_n  \int \limits_{nT_c}^{(n+1)T_c} p_c(t-nT_c)dt, \label{xi1b}
 \end{alignat}
\begin{alignat}{3}
\mathbf{x}_{i} &= \sqrt{\gamma_3 P_s} \frac{ \mathbf{h}}{N_cT_c} \sum_{i=1}^{N_s} \sum_{n=\frac{N_c}{N_s}(i-1)}^{\frac{N_c}{N_s}i-1} c_n s_i T_c, \label{xi1c}\\
&= \sqrt{\gamma_3 P_s} \frac{ \mathbf{h}}{N_c} \sum_{i=1}^{N_s} \sum_{n=\frac{N_c}{N_s}(i-1)}^{\frac{N_c}{N_s}i-1} c_n s_i,  \label{xi1d}
\end{alignat}
\end{subequations}}
\noindent where again the integration in~\eqref{xi1a} becomes the summation in~\eqref{xi1c} as mentioned above.
\noindent Substituting~\eqref{xs1ed} and~\eqref{xi1d} into~\eqref{rER}, we get~\eqref{rERapprox} which simplifies to~\eqref{rERapprox2} since ${\mathbf{f}}^T \mathbf{f}^* = {\mathbf{f}}^H \mathbf{f}$ and ${\mathbf{f}}^T \mathbf{h}^* = {\mathbf{f}}^H \mathbf{h}$ as given at the top of the page. From~\eqref{rERapprox2} the incident RF power on the ER can be found using~\eqref{qapprox} (also given at the top of the page).\\
\vspace{5mm}
\begin{figure*}[t]
\begin{align}\label{rERapprox}
{r}_{\textrm{ER}} &= \sqrt{\gamma_2 P_t} \frac{(\frac{\sqrt{\gamma_1 \gamma_2 P_s} g^*}{N_s} \sum_{i=1}^{N_s} s_i^* {\textbf{f}}^T \mathbf{f}^* + \frac{\sqrt{\gamma_3 P_s}}{N_c} \sum_{i=1}^{N_s} \sum_{n=\frac{N_c}{N_s}(i-1)+1}^{\frac{N_c}{N_s}i} c_n s_i^* {\textbf{f}}^T \mathbf{h}^* + \mathbf{f}^T {\mathbf{n}}^* )}{\left\|\frac{\sqrt{\gamma_1 \gamma_2 P_s} g^*}{N_s} \sum_{i=1}^{N_s} s_i {\textbf{f}}+ \frac{\sqrt{\gamma_3 P_s}}{N_c} \sum_{i=1}^{N_s} \sum_{n=\frac{N_c}{N_s}(i-1)+1}^{\frac{N_c}{N_s}i} c_n s_i  {\textbf{h}} + {\mathbf{n}}\right\|}.
\end{align}
\begin{align}\label{rERapprox2}
{r}_{\textrm{ER}} &= \sqrt{\gamma_2 P_t} \frac{(\frac{\sqrt{\gamma_1 \gamma_2 P_s} g}{N_s} \sum_{i=1}^{N_s} s_i^* {\textbf{f}}^H \textbf{f} + \frac{\sqrt{\gamma_3 P_s}}{N_c} \sum_{i=1}^{N_s} \sum_{n=\frac{N_c}{N_s}(i-1)+1}^{\frac{N_c}{N_s}i} c_n s_i^* {\textbf{f}}^H \textbf{h} + \mathbf{f}^H {\mathbf{n}} )}{\left\|\frac{\sqrt{\gamma_1 \gamma_2 P_s} g}{N_s} \sum_{i=1}^{N_s} s_i {\textbf{f}}+ \frac{\sqrt{\gamma_3 P_s}}{N_c} \sum_{i=1}^{N_s} \sum_{n=\frac{N_c}{N_s}(i-1)+1}^{\frac{N_c}{N_s}i} c_n s_i  {\textbf{h}} + {\mathbf{n}}\right\|}.
\end{align}
\begin{align}\label{qapprox}
Q_{RF} &= |r_{\textrm{ER}}|^2 \nonumber\\
&=  \left({\frac{\frac{\gamma_1 {\gamma_2}^2 P_s P_t |g|^2}{N_s^2} \left|\sum_{i=1}^{N_s} s_i\right|^2 {\left\| \mathbf{f} \right\|}^4  + \frac{\gamma_3 P_s P_t}{N_c^2} \left|\sum_{i=1}^{N_s} \sum_{n=\frac{N_c}{N_s}(i-1)+1}^{\frac{N_c}{N_s}i} c_n s_i\right|^2 {\left\|{\textbf{f}}^H \textbf{h}\right\|}^2 + \gamma_2 P_t {\left\|\mathbf{f}^H {\mathbf{n}}\right\|}^2 } {\frac{{\gamma_1 \gamma_2 P_s} |g|^2}{N_s^2} \left|\sum_{i=1}^{N_s} s_i^*\right|^2  {\left\| \mathbf{f} \right\|}^2 + \frac{\gamma_3 P_s}{N_c^2} \left|\sum_{i=1}^{N_s} \sum_{n=\frac{N_c}{N_s}(i-1)+1}^{\frac{N_c}{N_s}i} c_n s_i^*\right|^2  {\left\| \mathbf{h} \right\|}^2 +  \|{\mathbf{n}}\|^2 }}\right).
\end{align}
\rule{18.2cm}{0.5pt}
\vspace{-5mm}
\end{figure*}
Let
\begin{align}\label{mu_nu}
\mu &= \left|\sum_{i=1}^{N_s} s_i\right|^2 = \left|\sum_{i=1}^{N_s} s_i^*\right|^2,\nonumber\\
\nu &= \left|\sum_{i=1}^{N_s} \sum_{n=\frac{N_c}{N_s}(i-1)}^{\frac{N_c}{N_s}i-1} c_n s_i^*\right|^2 = \left|\sum_{i=1}^{N_s} \sum_{n=\frac{N_c}{N_s}(i-1)}^{\frac{N_c}{N_s}i-1} c_n s_i\right|^2.
\end{align}
\indent Asymptotic massive MIMO expressions for Rayleigh fading channels have been presented in~\cite{Lim-2015}. Following a similar procedure for Nakagami-$m$ fading channels, we can show that $\frac{1}{M} {\left\|\mathbf{f}_i\right\|}^4 \,\to\, M +\frac{1}{m_f} $, $\frac{1}{M} {\left\|\mathbf{f}_i\right\|}^2 \,\to\, 1 $, $\frac{1}{M} {\left\|{\mathbf{f}_k}^H  \mathbf{f}_i \right\|}^2 \,\to\, 1 $, $\frac{1}{M} {\mathbf{f}_k}^H  \mathbf{f}_i  \,\to\, 0 $ (for $k \neq i$),  $\frac{1}{M} {\mathbf{f}_k}^H  \widetilde {\mathbf{n}}  \,\to\, 0 $, $\frac{1}{M} {\left\|{\mathbf{f}_i}^H\widetilde {\mathbf{n}} \right\|}^2 \,\to\, \frac{{\sigma_n}^2}{NT_c} $ and $\frac{1}{M} {\left\|{\widetilde {\mathbf{n}}} \right\|}^2 \,\to\, \frac{{\sigma_n}^2}{NT_c} $.
Note that only the expression for ${\left\|\mathbf{f}_i\right\|}^4$ is different for Nakagami-$m$ channels as compared to Rayleigh fading, while the others remain the same. Also, only $m_f$ appears in the expression and $m_g$ and $m_h$ do not impact the results. Substituting these asymptotic results in~\eqref{qapprox} gives us the result in~\eqref{q_instant} for $N_s \leq N_c$ and is reproduced below
\sugcom{\begin{align*}
Q \approx \gamma_2 P_t \left(\frac{\gamma_1 {\gamma_2}  |g|^2 \mu \left(M+\dfrac{1}{m_f}\right) + \gamma_3 \nu \left(\frac{N_s}{N_c}\right)^2 + \frac{\sigma_n^2 N_s}{T_s P_s}} {\gamma_1 \gamma_2 |g|^2 \mu +  \gamma_3 \nu \left(\frac{N_s}{N_c}\right)^2  + \frac{\sigma_n^2 N_s}{T_s P_s}}\right).
\end{align*}}
When $N_s = N_c$,~\eqref{q_instant} simplifies to
\begin{align*}
Q & \approx \gamma_2 P_t \left(\frac{ \gamma_1 {\gamma_2} |g|^2 \mu \left(M+\dfrac{1}{m_f}\right) + \gamma_3 \nu +  \frac{\sigma_n^2 N_s}{T_s P_s}} {\gamma_1 \gamma_2 |g|^2 \mu + \gamma_3 \nu + \frac{\sigma_n^2 N_s}{T_s P_s}}\right).
\end{align*}
\underline{Case 2: $N_s \geq N_c$}
In this case, $T_s < T_c$. Substituting~\eqref{ambient} in \eqref{xs}, we have
{\small \begin{subequations}
\begin{alignat}{3}
\mathbf{x}_{s}&= \frac{\sqrt{\gamma_1 \gamma_2 P_s} g^* \mathbf{f}^*}{N_cT_c} \int \limits_{0}^{N_cT_c} \sum_{n=1}^{N_c}c_n^2 {p_c}^2(t-nT_c) \sum_{i=1}^{N_s} s_i p_{s}(t-iT_s)dt, \label{xs2a}\\
&= \sqrt{\gamma_1 \gamma_2 P_s} \frac{ g^* \mathbf{f}^*}{\ N_cT_c } \sum_{n=1}^{N_c} \sum_{i=\frac{N_s}{N_c}(n-1)}^{\frac{N_s}{N_c}n-1} s_i  \int \limits_{iT_s}^{(i+1)T_s}p_{s}(t-iT_s)dt, \label{xs2b}
\end{alignat}
\begin{alignat}{3}
&= \sqrt{\gamma_1 \gamma_2 P_s} \frac{ g^* \mathbf{f}^*}{\ N_cT_c } \sum_{n=1}^{N_c} \sum_{i=\frac{N_s}{N_c}(n-1)}^{\frac{N_s}{N_c}n-1} s_i T_s, \label{xs2c}\\
&= \sqrt{\gamma_1 \gamma_2 P_s}  \frac{g^* \mathbf{f}^*}{N_s} \sum_{i=1}^{N_s} s_i, \label{xs2d}
\end{alignat}
\end{subequations}}
where the~\eqref{xs2c} comes from the fact that $\int_{iT_s}^{(i+1)T_s}p_{s}(t-iT_s)dt = T_s$ and~\eqref{xs2d} is obtained using $N_cT_c = N_sT_s$.
\noindent Substituting~\eqref{ambient} in~\eqref{xi}, we obtain
{\small \begin{subequations}
\begin{alignat}{3}
\mathbf{x}_{i} &=  \frac{\sqrt{\gamma_3 P_s} \textbf{h}^H}{ N_cT_c} \int \limits_{0}^{N_cT_c} \sum_{i=1}^{N_s} s_i p_{s}(t-iT_s) \sum_{n=1}^{N_c} c_n p_c(t-nT_c)dt, \label{xi2a}\\
&= \sqrt{\gamma_3 P_s} \frac{ \textbf{h}^H}{N_cT_c} \sum_{n=1}^{N_c} \sum_{i=\frac{N_s}{N_c}(n-1)}^{\frac{N_s}{N_c}n-1} c_n s_i \int \limits_{iT_s}^{(i+1)T_s}p_{s}(t-iT_s)dt, \label{xi2b}
\end{alignat}
\begin{alignat}{3}
\mathbf{x}_{i}&= \sqrt{\gamma_3 P_s} \frac{ \textbf{h}^H}{N_cT_c} \sum_{n=1}^{N_c} \sum_{i=\frac{N_s}{N_c}(n-1)}^{\frac{N_s}{N_c}n-1} c_n s_i T_s, \label{xi2c}\\
&= \sqrt{\gamma_3 P_s} \frac{ \textbf{h}^H}{N_s} \sum_{n=1}^{N_c} \sum_{i=\frac{N_s}{N_c}(n-1)}^{\frac{N_s}{N_c}n-1} c_n s_i, \label{xi2d}
\end{alignat}
\end{subequations}}
\noindent where~\eqref{xi2c} and~\eqref{xi2d} follow from the same reasoning as in~\eqref{xs2c} and~\eqref{xs2d}.\\
\indent Substituting~\eqref{xs2d} and~\eqref{xi2d} into~\eqref{rER}, we get~\eqref{rER2}, from which the incident RF power can be found as given in~\eqref{q_it3} at the top of the page.
\begin{figure*}[t]
\begin{align}\label{rER2}
{r}_{\textrm{ER}} &= \sqrt{\gamma_2 P_t} \frac{( \frac{\sqrt{\gamma_1 \gamma_2 P_s} g}{N_s} \sum_{i=1}^{N_s} s_i {\textbf{f}}^H \textbf{f} +\frac{\sqrt{\gamma_3 P_s}}{N_s} \sum_{n=1}^{N_c} \sum_{i=\frac{N_s}{N_c}(n-1)+1}^{\frac{N_s}{N_c}n} c_n s_i {\textbf{f}}^H \textbf{h} + \textbf{f}^H  {\mathbf{n}})}{\left\|\frac{\sqrt{\gamma_1 \gamma_2 P_s} g^*}{N_s} \sum_{i=1}^{N_s} s_i^* {\textbf{f}}+  \frac{\sqrt{\gamma_3 P_s}}{N_s} \sum_{n=1}^{N_c} \sum_{i=\frac{N_s}{N_c}(n-1)+1}^{\frac{N_s}{N_c}n} c_n s_i^*  {\textbf{h}} +  {\mathbf{n}}\right\|}.
\end{align}
\begin{align}\label{q_it3}
Q_{RF} &= |r_{\textrm{ER}}|^2 \approx \left({\frac{ \frac{\gamma_1 {\gamma_2}^2 P_s P_t |g|^2}{N_s^2} \left|\sum_{i=1}^{N_s} s_i\right|^2 {\left\| \mathbf{f} \right\|}^4  + \frac{\gamma_3 P_s P_t }{N_s^2} \left|\sum_{n=1}^{N_c} \sum_{i=\frac{N_s}{N_c}(n-1)+1}^{\frac{N_s}{N_c}n} c_n s_i\right|^2 {\left\|{\textbf{f}}^H \textbf{h}\right\|}^2 + \gamma_2 P_t {\left\|\textbf{f}^H  {\mathbf{n}}\right\|}^2 } { \frac{{\gamma_1 \gamma_2 P_s} |g|^2}{N_s^2} \left|\sum_{i=1}^{N_s} s_i^*\right|^2  {\left\| \mathbf{f} \right\|}^2 + \gamma_3 P_s \frac{1}{N_s^2} \left|\sum_{n=1}^{N_c} \sum_{i=\frac{N_s}{N_c}(n-1)+1}^{\frac{N_s}{N_c}n} c_n s_i^*\right|^2  {\left\| \mathbf{h} \right\|}^2 +  \| {\mathbf{n}}\|^2 }}\right)
\end{align}
\rule{18.2cm}{0.5pt}
\vspace{-5mm}
\end{figure*}
\eqref{q_it3} when simplified using the asymptotic massive MIMO expressions~\cite{Lim-2015}, gives the result for $N_s \geq N_c$ in~\eqref{q_instant}, reproduced below:
\begin{align*}
Q \approx \gamma_2 P_t \left(\frac{\gamma_1 {\gamma_2}  |g|^2 \mu \left(M+\dfrac{1}{m_f}\right) + \gamma_3 \nu + \frac{\sigma_n^2  N_s}{T_s P_s}} {\gamma_1 \gamma_2 |g|^2 \mu + \gamma_3 \nu  + \frac{\sigma_n^2  N_s}{T_s P_s}}\right),
\end{align*}
where $\mu $ and $\nu$ are as defined in~\eqref{mu_nu}.
\end{appendices}


\ifCLASSOPTIONcaptionsoff
\fi


\end{document}